\documentclass[aps,twocolumn,showpacs,floats,amssymb,prl]{revtex4-1}

\usepackage{graphicx}
\usepackage{dcolumn}
\usepackage{bm}
\usepackage{color}
\usepackage{amsmath}
\usepackage{epstopdf}
\usepackage{hyperref}
\usepackage{xcolor}
\usepackage{ulem}
\usepackage{footnote}

\begin{document}

\author{L. A. Pe\~{n}a Ardila$^{1,2}$, G. E. Astrakharchik$^{3}$ and S. Giorgini$^{4}$}

\affiliation{$^{1}$Institut f\"ur Theoretische Physik, Leibniz Universit\"at, 30167 Hannover, Germany}
\email{luis.ardila@itp.uni-hannover.de}
\affiliation{$^{2}$Institut for Fysik og Astronomi, Aarhus Universitet, 8000 Aarhus C, Denmark}
\affiliation{$^{3}$Departament de F\'{i}sica, Universitat Polit\`{e}cnica de Catalunya,
Campus Nord B4-B5, E-08034, Barcelona, Spain}
\affiliation{$^{4}$Dipartimento di Fisica, Universit\`a di Trento and CNR-INO BEC Center, I-38123 Povo, Trento, Italy}

\title{Strong coupling Bose polarons in a two-dimensional gas}

\begin{abstract} 
We study the properties of Bose polarons in two dimensions using quantum Monte Carlo techniques. Results for the binding energy, the effective mass and the quasiparticle residue are reported for a typical strength of interactions in the gas and for a wide range of impurity-gas coupling strengths. A lower and an upper branch of the quasiparticle exist. The lower branch corresponds to an attractive polaron and spans from the regime of weak coupling, where the impurity acts as a small density perturbation of the surrounding medium, to deep bound states which involve many particles from the bath and extend as far as the healing length. The upper branch corresponds to an excited state where due to repulsion a low density bubble forms around the impurity, but might be unstable against decay into many-body bound states. Interaction effects strongly affect the quasiparticle properties of the polaron. In particular, in the strongly correlated regime the impurity features a vanishing quasiparticle residue, signalling the transition from an almost free quasiparticle to a bound state involving many atoms from the bath.
\end{abstract}

\date{\today}

\maketitle

\section{I. Introduction}
Impurities embedded in a quantum many-body environment can lead to the formation of quasiparticles coined polarons. The concept was first introduced by Landau and Pekar in the solid-state context to describe an electron coupled to an ionic crystal~\cite{Landau48}. Polarons are fundamental ingredients in many different transport phenomena across condensed-matter physics. Electronic transport in polar crystals or semiconductors~\cite{PolaronsandExcitonsBook} as well as charge and spin transport in organic materials~\cite{Gershenson78,Watanabe14} can be understood in terms of polarons. Pairing between polarons is relevant in the physics of high-temperature superconductors~\cite{Lee06} and polarons are also candidates for electronic transport in DNA and proteins~\cite{EffectiveModelsforChargeTransportinDNANanowires}. Furthermore, polarons are used as probes of quantum many-body systems. For example, the low-energy excitations in a strongly correlated superfluid such as $^{4}$He can be probed by $^{3}$He impurity atoms~\cite{LandauFermiLiquidTheory}.

\begin{figure}[!h]
\begin{center}
\includegraphics[width=1\linewidth]{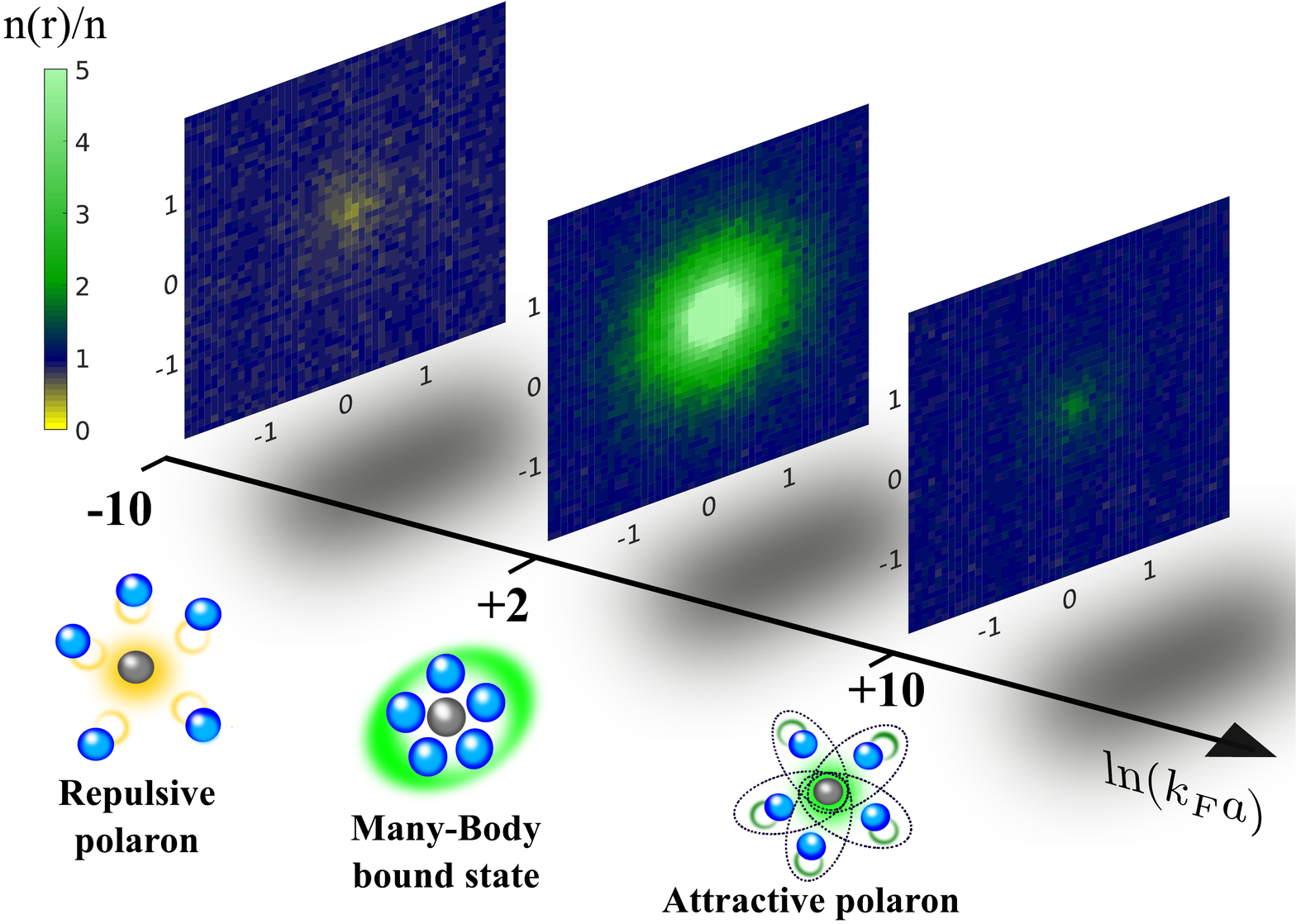}
\end{center}
\caption{Average over many snapshots of the particle positions around the impurity for three characteristic values of the impurity-bath coupling constant $\ln(k_{F}a)$. The impurity is located in the center and distances are in units of the healing length $\xi$. The local density $n(r)$ is estimated by summing particles over a square grid of size $L/100$, where $L$ is the size of the simulation box, and the color-bar indicates the ratio $n(r)/n$ over the bulk density $n$. At weak coupling impurities form polarons, {\it{i.e.}} almost free quasiparticles slightly dressed by the medium (right and left upper panels). On the attractive branch as the coupling $\ln(k_Fa)>0$ is decreased, the impurity forms a many-body bound state involving up to few tens of particles and whose size is as large as the healing length (central upper panel).}
\label{M}
\end{figure}

Unprecedented control and versatility of ultracold gases~\cite{Bloch08} made it possible to experimentally observe dressed impurities named Fermi and Bose polarons depending on whether they interact, respectively, with a degenerate Fermi gas~\cite{Schirotzek09, Koschorreck12, Kohstall12, Cetina16, Scazza17} or a Bose-Einstein condensate (BEC). For Bose polarons, experiments have been carried out in three-dimensional (3D)~\cite{Jorgensen16,Hu16,Camargo18,Ardila18,Zwierlein19} and one-dimensional (1D)~\cite{Spethmann12,Catani12} geometries. Observation of Bose polarons in ultracold gases has triggered an intense research activity aiming at describing the crossover from weak to strong coupling regimes. In the former case the so-called Bogoliubov-Fr\"ohlich Hamiltonian describes accurately the ground state properties of the polaron~\cite{Viverit02, Tempere09, Grusdt15, Grusdt15_2, Kain14, Vlietinck15, Levinsen17, Lampo2018, Nielsen18, DipolarPolaron18}. However, quantum fluctuations become relevant as interactions are increased, making the description in terms of the Fr\"ohlich paradigm inadequate. By using the Gross-Pitaevskii equation strongly interacting Bose polarons were predicted to manifest exotic phenomena such as self-localization~\cite{Cucchietti06,Kalas06, Bruderer08, Santamore11, Blinova13}, but without experimental evidence  so far. Recently, properties of these strongly coupled impurities have also been addressed by techniques such as $T$-matrix, diagrammatic and variational approaches which go beyond the single-phonon excitation scheme of the Fr\"ohlich model~\cite{Rath13, Li14,Christensen15,Ardila15,Shchadilova16,Grusdt2017, Kain18}. These studies predict exotic out of equilibrium dynamics, non trivial quasiparticle splitting due to finite-temperature effects~\cite{Abdelaali14,Abdelaali14-2,Mistakidis18,Mistakidis19,Shchadilova16,Drescher18,Guenther18} as well as important few-body effects~\cite{Levinsen15,Sun17}. Furthermore, the regime of strong coupling should also feature the interchange of Bogoliubov modes between polarons via polaron-polaron interactions~\cite{Dehkharghani18, Camacho2018b}. In the context of theoretical techniques suitable to investigate this latter regime, the quantum Monte-Carlo (QMC) method is based on a microscopic Hamiltonian and provides {\it exact} (within controllable statistical errors) ground-state properties of the polaron for arbitrary coupling strengths~\cite{Ardila15,Ardila16,Parisi16, Grusdt2017}.

Physically, the two-dimensional (2D) geometry is appealing since the role of quantum fluctuations is enhanced while off-diagonal long-range order, responsible of BEC phenomena, still exists in the ground state. Polarons in 2D geometries have been extensively investigated in the context of Fermi polarons~\cite{Koschorreck12,Massignan14,Nga12,Nga13,SchmidtFermi} and exciton impurities coupled to semiconductors~\cite{Sidler17}. Bose polarons have been investigated within the context of the Fr\"ohlich model~\cite{Casteels12,Grusdt16,Pastukhov18}, but a quantitatively precise description of 2D Bose polarons in the strongly coupled regime is still lacking.

Here, we use exact QMC methods~\cite{Ardila15, Ardila16,Parisi16} to study an impurity immersed in a 2D Bose superfluid and to compute the polaron energy, the effective mass and the  quasiparticle residue for arbitrary coupling strength. Quantitatively significant deviations of the quasiparticle properties from perturbation theory are found already at weak coupling strengths between the impurity and the bath. In the strongly interacting regime, the polaron loses the quasiparticle nature characteristic of weak interactions: the wavefunction residue vanishes indicating that the coherence is lost. In this regime the impurity is not longer free to move, but instead is bound to a density perturbation which involves many particles from the bath (see Fig.~\ref{M}).
\section{II. SYSTEM AND PERTUBATION THEORY} We consider an impurity of mass $m_{I}$ embedded in a 2D Bose gas consisting of $N$ atoms of mass $m_{B}$ at $T=0$ in a square box of size $L$ with overall density $n=\frac{N}{L^2}$. In the first quantization formalism the Hamiltonian of the system reads
\begin{eqnarray}
H&=&-\frac{\hbar^2}{2m_B}\sum_{i=1}^N \nabla_i^2+\sum_{i<j}V_B(r_{ij}) 
\nonumber\\
&-&\frac{\hbar^2}{2m_I}\nabla_\alpha^2+\sum_{i=1}^NV_I(r_{i\alpha}) \;.
\label{Hamiltonian}
\end{eqnarray}
Here, the first two terms represent the kinetic and the interaction energy of the bosonic bath where particles interact through the two-body potential $V_B$, which depends on the distance $r_{ij}=|{\bf r}_i-{\bf r}_j|$ between a pair of bosons. Furthermore, $-\frac{\hbar^2\nabla_\alpha^2}{2m_I}$ is the kinetic energy of the impurity denoted by the coordinate vector 
${\bf r}_\alpha$ and $V_I$ is the boson-impurity potential depending on the distance $r_{i\alpha}=|{\bf r}_\alpha-{\bf r}_i|$ between the impurity and the $i$-th bath particle. Both interaction potentials $V_B$ and $V_I$ are short ranged and are parameterized by the scattering lengths $a_B$ and $a$, respectively.
Within Bogoliubov theory, the Hamiltonian (\ref{Hamiltonian}) can be written in second quantization as the sum of two terms 
$H=H_0+H_{\text{int}}$, where 
\begin{equation}
H_0=\frac{p^2}{2m_I}+E_B+\sum_{\bf k} \epsilon_k \alpha_{\bf k}^\dagger\alpha_{\bf k} \;,
\label{Hamiltonian0}
\end{equation}
is the unperturbed Hamiltonian of a free impurity moving with momentum ${\bf p}$ and a static host gas. The bath is described in terms of non-interacting Bogoliubov excitations with energy $\epsilon_k=\sqrt{\left(\epsilon_k^0\right)^{2}+2g_Bn\epsilon_k^0}$, where $\epsilon_k^0=\frac{\hbar^2k^2}{2m_B}$ is the dispersion of free particles and $g_B=\frac{4\pi\hbar^2/m_B}{\ln(1/na_B^2)}$ is the 2D density-dependent coupling constant of the Bose gas. The ground state of the bath corresponds to the vacuum of excitations and has energy $E_B$. The interaction Hamiltonian $H_{\text{int}}$ is given by the sum of a mean-field shift and a term where the impurity is coupled to the creation and annihilation operators of single excitations in the Bose gas
\begin{equation}
H_{\text{int}}=gn+\frac{g\sqrt{n}}{\sqrt{L^{2}}}\sum_{\bf q}e^{i\bf{q\cdot r_\alpha}}\sqrt{\frac{\epsilon_q^0}{\epsilon_q}}\left(\alpha_{\bf q}+\alpha_{-{\bf q}}^{\dagger}\right)\;.
\label{Hamiltonian1}
\end{equation}
Here, $g=\frac{2\pi\hbar^{2}/m_r}{\ln\left(1/na^{2}\right)}$ is the 2D effective coupling constant which contains the reduced mass $m_r=\frac{m_I m_B}{m_I+m_B}$. It describes the scattering processes between the impurity and the bath particles in terms of the scattering length $a$ of the potential $V_I$. The above Hamiltonian $H_0+H_{\text{int}}$ embodies the well-known Fr\"ohlich model which is expected to correctly describe the physics of Bose polarons in the weakly interacting limit, where coupling to multiple excitations of the bath can be neglected~\cite{Levinsen15}.

If $E({\bf p})$  is the energy of the impurity-bath system where the impurity has momentum ${\bf p}$, the low momentum expansion of the energy difference
\begin{equation}
E({\bf p})-E_B=\mu+\frac{p^2}{2m_I^\ast}+\dots \;,
\label{energy}
\end{equation}
defines the binding energy $\mu$ of the impurity and its effective mass $m_I^\ast$. By using perturbation theory one finds the following results holding for $m_I=m_B=m$ to lowest order in the coupling strength $g$ of the interaction Hamiltonian $H_{\text{int}}$ (see Appendix)
\begin{equation}
\frac{\mu}{\mu_{0}}=\frac{4}{\ln\left(4\pi\right)-2\ln\left(k_{F}a\right)},
\label{penergy}
\end{equation}
and
\begin{equation}
\frac{m}{m^\ast}=1-\frac{1}{2}\frac{\left[\ln\left(4\pi\right)-2\ln\left(k_{F}a_{B}\right)\right]}{\left[\ln\left(4\pi\right)-2\ln\left(k_{F}a\right)\right]^{2}}.
\label{pmass}
\end{equation}
Here we use $\mu_{0}=\frac{\hbar^{2}k^{2}_{F}}{2m}$, involving the Fermi wavevector $k_{F}=\sqrt{4\pi n}$ of a system having the same density $n$ of the gas. Moreover, the coupling strength of the impurity-bath interaction is expressed in terms of $\ln(k_{F}a)$. These results were first derived in Ref.~\cite{Pastukhov18}. The same perturbation approach allows one to calculate the overlap $\sqrt{Z_p}$ between the interacting and non-interacting ground state of the impurity-bath system with the impurity moving with momentum ${\bf p}$. For the impurity at rest (${\bf p}=0$) one finds~\cite{Pastukhov18} (see Appendix)
\begin{equation}
Z_{0}=1-\frac{\left[\ln\left(4\pi\right)-2\ln\left(k_{F}a_{B}\right)\right]}{\left[\ln\left(4\pi\right)-2\ln\left(k_{F}a\right)\right]^{2}}.
\label{presidue}
\end{equation}
Notice that the above results hold in the weak-coupling regime $|\ln(k_Fa)|\gg1$ of the impurity-bath interaction.

\section{III. QMC RESULTS} 
In order to calculate the properties of the polaron for all coupling strengths we resort to QMC techniques. Details on the general method can be found in Refs.~\cite{ThesisPOL,Ardila15,Ardila16}, whereas an exhaustive discussion of the interatomic potentials used in the simulations and on the trial wavefunction used for importance sampling are found in the Appendix. Simulations are performed for a gas of $N$ identical particles and a single impurity in a square box of size $L$ with periodic boundary conditions. We choose the value $\tilde{g}_B=\frac{m_Bg_B}{\hbar^2}=0.136$ for the dimensionless coupling constant in the bath. This value corresponds to $|\ln(k_Fa_B)|\simeq45$ and is typical for the experimental conditions of 2D Bose gases\cite{Ville18}. Furthermore, as in the perturbation theory study, we consider the case where impurity and particles in the bath have the same mass: $m_B=m_I=m$. 
\begin{figure}
\begin{center}
\includegraphics[width=1\linewidth]{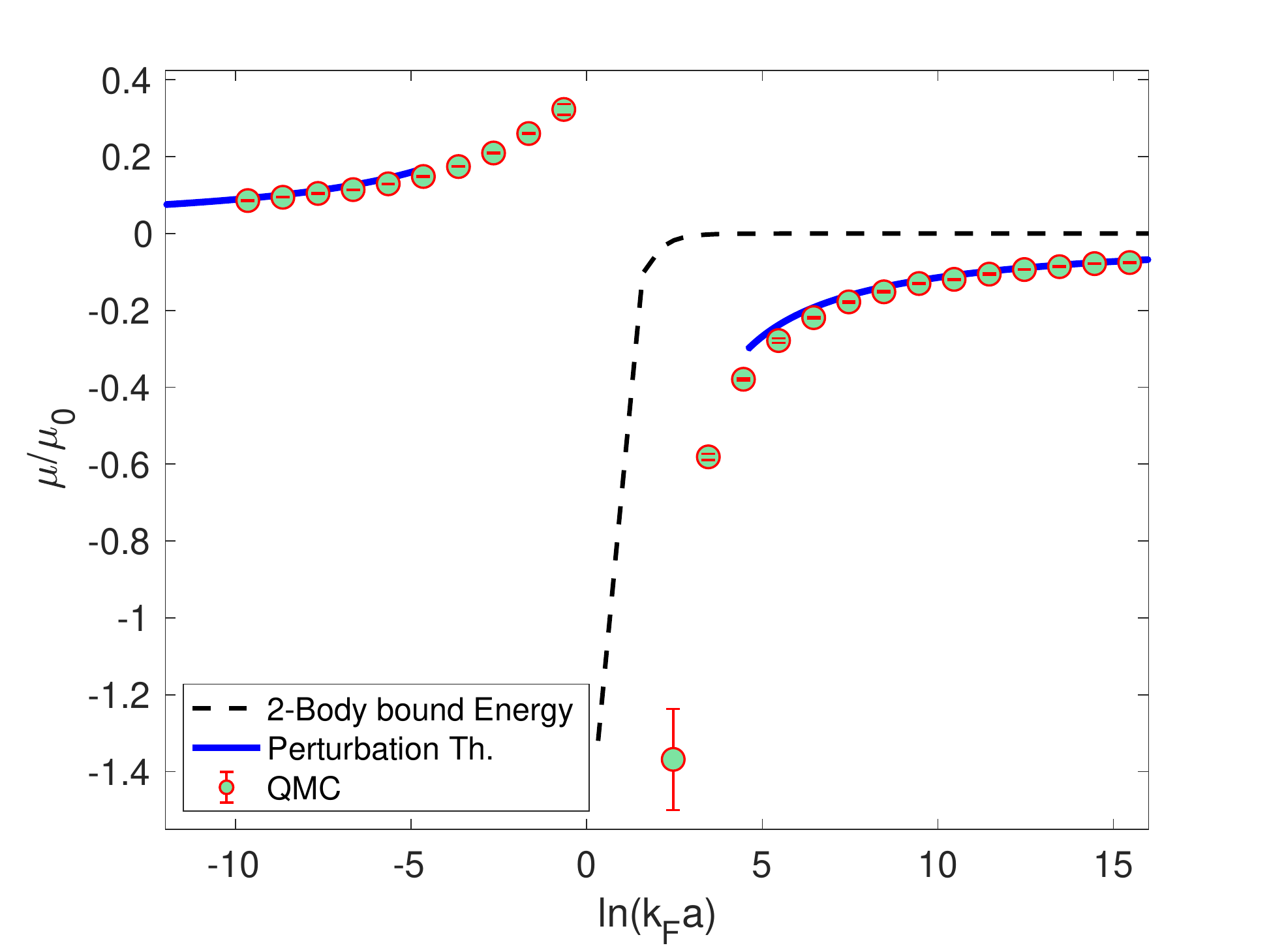}
\caption{Polaron energy as a function of the coupling strength $\ln(k_{F}a)$ for the attractive and repulsive branches (circles). The dashed line shows the dimer binding energy $\epsilon_b$ and the solid line the perturbation result from Eq.~(\ref{penergy}). The coupling constant of the gas is $\tilde{g}_B=0.136$.}
\label{ENERGY}
\end{center}
\end{figure}

We calculate the polaron energy $\mu$ from the direct calculation of the ground-state energy of the bath with and without the impurity, $\mu=E(N,1)-E(N)$, where $N$ is the number of particles in the bath. Results are shown in Fig.~\ref{ENERGY}. In analogy with the 2D Fermi polaron~\cite{SchmidtFermi,Koschorreck12} we find two branches: one corresponds to the ground state of the attractive polaron with $\mu<0$ and the second to an excited state of the quasiparticle with $\mu>0$. It is important to notice that a two-body bound state with energy $\epsilon_b$ exists for any value of the coupling constant $\ln{(k_Fa)}$. This is in contrast with the 3D polaron where the dimer state only appears as the $s$-wave scattering length turns positive, on one side of the scattering resonance~\cite{Ardila18}. In the weakly interacting regime, $|\ln(k_{F}a)|\gg1$, the QMC results are in good agreement with the prediction (\ref{penergy}) of perturbation theory. Following the attractive branch, we notice that the polaron energy is always much larger, in absolute value, than the dimer binding energy $\epsilon_b$. This is due to many-body effects which favour the formation of cluster states around the impurity involving many particles of the bath (see Fig.~\ref{M} central panel). In particular, in the vicinity of $\ln(k_{F}a)\approx 0$ large fluctuations occur in our QMC simulations due to the formation of very deep many-body bound states which makes both the attractive and the repulsive branch of the polaron hard to follow further. The repulsive branch describes an excited state where the impurity repels the particles of the bath at large distance, but is unstable against cluster formation at short distance. The state is well defined provided the typical size $a$ of bound states is small compared to the average interparticle distance $k_F^{-1}$, but gets increasingly ill defined as the two length scales become comparable. This is exactly what we observe in our simulations where the excited state of the polaron is described using an appropriate choice of the wavefunction used for importance sampling (see Appendix for more details).

\begin{figure}[!h]
\begin{center}
\includegraphics[width=1\linewidth]{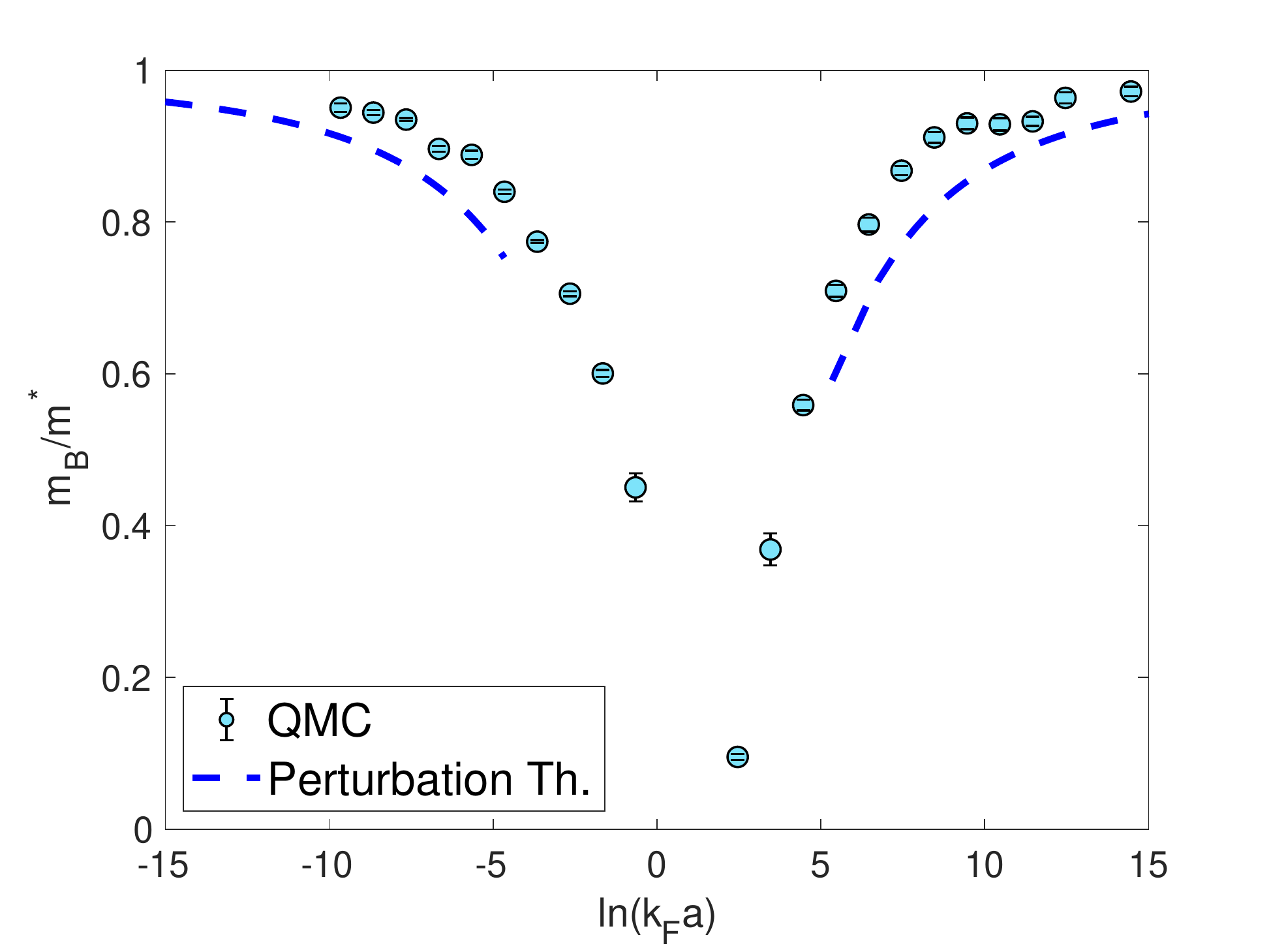}
\end{center}
\caption{Effective mass $m^\ast$ of the polaron as a function of the coupling strength $\ln(k_{F}a)$. The coupling constant of the Bose gas is $\tilde{g}_B=0.136$.}
\label{MASS}
\end{figure}

\begin{figure}
\begin{center}
\includegraphics[width=1\linewidth]{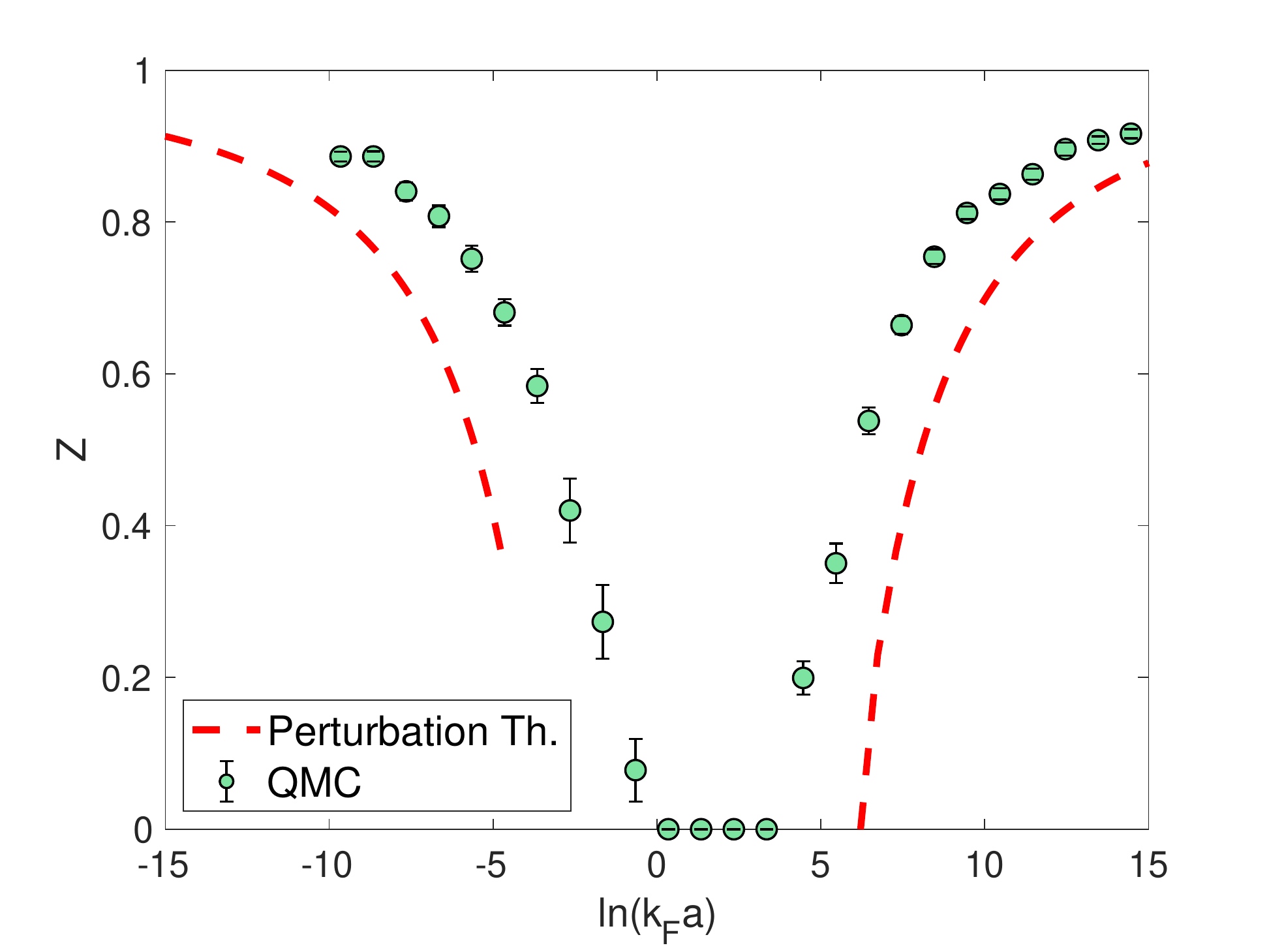}
\end{center}
\caption{Residue $Z_0$ of the polaron as a function of the coupling strength $\ln(k_{F}a)$~\cite{note1}. The coupling constant of the gas is $\tilde{g}_B=0.136$.}
\label{Z}
\end{figure}

Furthermore, we study the mobility of the impurity by calculating its effective mass $m^{\ast}$ as a function of the coupling strength $\ln{k_Fa}$. The effective mass is determined by computing the mean-square displacement of the impurity in imaginary time~\cite{Ardila15}
\begin{equation}
\frac{m}{m^\ast}=\lim_{\tau\rightarrow\infty}\frac{\left\langle \left|\Delta\mathbf{r}_{\alpha}(\tau)\right|^{2}\right\rangle }{4D\tau} \;,
\end{equation}
where $D=\hbar^2/(2m)$ is the diffusion constant of a free particle and $\langle \left|\Delta{\bf r}_{\alpha}(\tau)\right|^{2}\rangle =\langle \left|{\bf r}_{\alpha}(\tau)-{\bf r}_{\alpha}(0)\right|^{2}\rangle$, being $\tau=it/\hbar$ the imaginary time of the QMC simulation. The effective mass is found by fitting the slope of the mean-square displacement for large values of $\tau$. The  residue $Z_0$ of the polaron is obtained from the one-body density matrix associated to the impurity
\begin{equation}
\rho(\mathbf{r})=\left\langle \frac{\psi_{T}(\mathbf{r}_{\alpha}+\mathbf{r},\mathbf{r}_{1},\cdots,\mathbf{r}_{N})}{\psi_{T}(\mathbf{r}_\alpha,\mathbf{r}_{1},\cdots,\mathbf{r}_{N})}
\label{OBDM}
\right\rangle \;,
\end{equation}
where $\psi_T$ is the many-body guiding wavefunction of the QMC simulation. The above quantity is normalized to unity for ${\bf r}\to0$, whereas its long-range limit gives 
the residue
\begin{equation}
\lim_{r\rightarrow\infty}\rho(\mathbf{r})\rightarrow Z_0\;.
\end{equation}

In Figs.~\ref{MASS}-\ref{Z} we show the results for the effective mass and the quasiparticle residue respectively. The calculation of the residue $Z_0$ is particularly sensitive to finite-size effects which make the extrapolation to the thermodynamic limit delicate. We have chosen different long-range asymptotic behaviours for the Jastrow terms entering the trial wavefunction (see Appendix). In the bath the long-range decay of boson-boson correlations is governed by phonons as shown in Ref.~\cite{ReattoChester67}. For the impurity-boson correlations, instead, we use the same functional form as from the Gross-Pitaevskii equation in the case of a static impurity. This choice of the trial wavefunction exhibits a fast convergence of $Z_0$ with increasing system sizes and allows us to keep finite-size effects under control(see Appendix). From Figs.~\ref{MASS}-\ref{Z} we notice that, even for the smallest reported values of the coupling ($|\ln(k_Fa)|\simeq10$), perturbation theory does not reproduce the QMC results of $m/m^\ast$ and $Z_0$. This is in contrast with the results of the polaron energy reported in Fig.~\ref{ENERGY} and shows that higher order terms, not accounted for by the Fr\"ohlich model, play an important role for these quantities already at such large values of $|\ln(k_Fa)|$~\cite{note}. In the regime of strong interactions both the inverse effective mass and the quasiparticle residue become significantly smaller than the corresponding non-interacting values. Indeed, we find that the polaron looses its quasiparticle nature as it gets more dressed by the particles from the bath. The perturbation caused by the impurity in the surrounding medium involves up to few tens of particles over a distance on the order of the healing length. We find a vanishing quasiparticle residue and a large effective mass which signal the transition to a many-body bound 
state (cluster state) without breaking of translational symmetry (localization). A similar situation occurs for Fermi polarons~\cite{SchmidtFermi,Koschorreck12} where Pauli exclusion principle only allows for the formation of a molecular state involving just one particle from the bath. A question which remains open also in the Fermi polaron case is whether the quasiparticle to bound state transition is discontinuous or continuous. 

Our findings are also in contrast with Bose polarons in 3D and 1D. In
fact, polarons in 3D remain well defined quasiparticles up to the limit of 
resonant interactions ~\cite{Jorgensen16,Hu16}, while in 1D they are never well defined  quasiparticles as the one-body density matrix (\ref{OBDM}) decays to zero at large distances with algebraic law for any value of the coupling constant between the impurity and the bath. In this respect, 2D geometry is peculiar because the quasiparticle nature of polarons is rapidly suppressed by 
increasing the interaction strength.

\section{IV. EXPERIMENTAL IMPLEMENTATION}

In Ref.~\cite{Ville18} a gas of $^{87}$Rb atoms in the hyperfine state $\left|F=1,m=0\right\rangle$ is confined in a 2D rectangular box with dimensions $L_{x}\approx L_{y}\approx30$ $\mu$m, at temperatures much below the Berezinskii-Kosterlitz-Thouless critical temperature. In the transverse direction a strong harmonic confinement is applied with frequency $\omega_{z}/(2\pi)\simeq4.6$ kHz and by changing the number of trapped atoms the 2D density can be varied in the range $n\approx 10-80 \;\mu\text{m}^{-2}$. The 2D scattering length is given by $a_{B}=1.863\;\ell_{z}\exp\left(-\sqrt{\frac{\pi}{2}}\ell_{z}/a_B^{(3D)}\right)$ (see Ref.~\cite{Bloch08}), in terms of the 3D s-wave scattering length $a_B^{(3D)}$ and the transverse length $\ell_{z}=\sqrt{\hbar/m_{B}\omega_{z}}$. With the typical ratio of lengths $\ell_z/a_B^{(3D)}\simeq30-50$ reached in experiments the interaction strength is in the range $\tilde{g}_B\simeq0.10-0.16$, where $\tilde{g}_B$ is the dimensionless parameter $\tilde{g}_B=\frac{m_Bg_B}{\hbar^2}=\sqrt{8\pi}a_B^{(3D)}/\ell_z$. Due to the exponential dependence of $a_B$ on the ratio $\ell_z/a_B^{(3D)}$, the 2D gas parameter takes on very small values: $na_B^2\simeq10^{-30}-10^{-50}$. In our purely 2D simulations we use the value $na_B^2=10^{-40}$ for the gas parameter of the bath, which corresponds to the effective 2D coupling strength $\tilde{g}_B=\frac{4\pi}{\ln(1/na_B^2)}=0.136$ close to the experimental conditions of Ref.~\cite{Ville18}.

\section{V. CONCLUSIONS}
We investigated the properties of Bose polarons in two dimensions. The polaron energy, effective mass and  quasiparticle residue have been calculated using QMC techniques for arbitrary coupling strength. We study the properties of the attractive and repulsive branch which correspond to the ground state and to a metastable state of the impurity. In the ground state the polaron energy is much lower than the one of the two-body bound state, which in 2D is present for any value of the interaction strength. At stronger couplings, the impurity forms a bound state involving many particles from the bath, which features a large effective mass and a vanishing wavefunction residue. A vanishing quasiparticle residue and a large effective mass  signal the transition from a polaron to a many-body bound state without breaking of translational symmetry.  A similar behaviour is found along the repulsive branch where a low-density bubble is formed around the impurity. However, this state rapidly becomes unstable against cluster formation as the interaction strength is increased. Our study is important for the investigation of transport properties in layered structures of ultracold atoms~\cite{Chien15,Krinner17} as well as layered solid-state materials~\cite{Sidler17}. 

\section{ACKNOWLEDGEMENTS}

This research was funded by the DFG Excellence Cluster QuantumFrontiers. S.G acknowledges funding from the Provincia Autonoma di Trento. G. E. A. acknowledges funding from the Spanish MINECO (FIS2017-84114-C2-1-P). The Barcelona Supercomputing Center (The Spanish National Supercomputing Center - Centro Nacional de Supercomputaci\'{o}n) is acknowledged for the provided computational facilities (RES-FI-2019-2-0033).

\section{APPENDIX A: GROUND-STATE PROPERTIES-PERTURBATION THEORY}

The Fr\"ohlich Hamiltonian in Eq.~[3] of the main text reads,
\begin{equation}\tag{A1}
H=H_{0}+gn+\frac{g\sqrt{n}}{\sqrt{L^{2}}}\sum_{\mathbf{k}}\exp\left(i\mathbf{k\cdot r_{\alpha}}\right)\sqrt{\frac{\epsilon_{\mathbf{k}}^{0}}{\epsilon_{\mathbf{k}}}}(\alpha_{\mathbf{k}}+\alpha_{-\mathbf{k}}^{\dagger})\;,
\end{equation}
where $H_{0}$ is the unperturbed Hamiltonian in Eq.~[2], consisting of an impurity with momentum $\mathbf{p}$ and independent Bogoliubov excitations with energy $\epsilon_{\mathbf{k}}$. Hence, the unperturbed ground state is represented by $\left|0\right\rangle =\left|\mathbf{p},0_{\mathbf{k}}\right\rangle$ and corresponds to the free impurity and the vacuum of Bogoliubov excitations. Relevant
processes consist in states in which the impurity is scattered by a single excitation. These states are represented by  $\left|\mathbf{k}\right\rangle =\left|\mathbf{p-\hbar\mathbf{k}},1_{\mathbf{k}}\right\rangle$ and correspond to unperturbed energies $E_{\mathbf{k}}^{(0)}=\frac{(\mathbf{p}-\hbar\mathbf{k})^2}{2m_I}+E_B+\epsilon_{\mathbf{k}}$.

\textit{Polaron energy:} The Hamiltonian is split into $H=H_{0}+H_{\text{int}}$. The energy expansion within perturbation theory is written as $E_{0}=E_{0}^{(0)}+E_{0}^{(1)}+E_{0}^{(2)}+\mathellipsis$, where $E_{0}^{(0)}=\frac{p^2}{2m_I}+E_B$ is the ground-state energy of the unperturbed system. The first and second order contributions to the energy are given by
\begin{equation}\tag{A2}
    \begin{array}{c}
E_{0}^{(1)}=\left\langle0\right|H_{\text{int}}\left|0\right\rangle, \\
\\
E_{0}^{(2)}=\sum_{\mathbf{k}\neq 0}\frac{\left|\left\langle\mathbf{k}\right|H_{\text{int}}\left|0\right\rangle \right|^{2}}{E_{0}^{(0)}-E_{\mathbf{k}}^{(0)}}.
\end{array}
\label{B2}
\end{equation}
By computing the matrix element $E_{0}^{(1)}$ one straightforwardly obtains,
\begin{equation}\tag{A3}
E_{0}^{(1)}=gn=\frac{4\pi\hbar^{2}n}{m}\frac{1}{\ln\left(1/na^{2}\right)}\;,
\label{mu}
\end{equation}
where we assumed equal masses for impurity and particles in the bath ($m_B=m_I=m$). The polaron energy $\mu$ is estimated using the lowest order result~(\ref{mu}) which, 
in units of the energy $\mu_{0}$ and in terms of the wavevector $k_{F}$, is written as in Eq.[5] of the main text.\\

\textit{Effective mass:} We assume that the energy of the bath with the impurity is written as $E(\mathbf{p})=E_{B}+\mu+\frac{p^{2}}{2m^\ast}+\mathellipsis$, holding at low momenta $\mathbf{p}$ of the impurity. Thus, in terms of the second order energy correction, the polaron mass is renormalized as
\begin{equation}\tag{A4}
\frac{1}{m^\ast}= \frac{1}{m}+\lim_{\mathbf{p}\to0}\frac{2\left(E_{0}^{(2)}-\mu+gn\right)}{p^2}\;.
\label{effectivemassdef}
\end{equation}
The term $E_{0}^{(2)}$ is given by
\begin{equation*}
E_{0}^{(2)}=-g^{2}\frac{n}{L^{2}}\sum_{\mathbf{k}}\left[\frac{\left(\mathbf{k}\xi\right)^{2}}{\left(\mathbf{k}\xi\right)^{2}+2}\right]^{1/2}\frac{1}{\Omega(\mathbf{k})}\;,
\end{equation*}
where we use $\frac{\epsilon_{\mathbf{k}}^{0}}{\epsilon_{\mathbf{k}}}=\left[\frac{\left(\mathbf{k}\xi\right)^{2}}{\left(\mathbf{k}\xi\right)^{2}+2}\right]^{1/2}$ with $\xi=\hbar/\sqrt{2mg_Bn}$ 
the healing length in the bath. In addition, we introduce the quantity $\Omega(\mathbf{k})=\frac{\left(\mathbf{p}-\hbar\mathbf{k}\right)^{2}}{2m}+\epsilon_{\mathbf{k}}-\frac{\mathbf{p}^{2}}{2m}$. At low momenta, $\mathbf{p}\rightarrow0$, one can expand $1/\Omega(\mathbf{k})$ as
\begin{equation}\tag{A5}
\frac{1}{\Omega(\mathbf{k})}\simeq\frac{1}{\frac{\hbar^{2}k^{2}}{2m}+\epsilon_{\mathbf{k}}}+\frac{\hbar\mathbf{k\cdot p}/m}{\frac{\hbar^{2}k^{2}}{2m}+\epsilon_{\mathbf{k}}}+\frac{\hbar^{2}k^{2}p^{2}/m^2}{\left(\frac{\hbar^{2}k^{2}}{2m}+\epsilon_{\mathbf{k}}\right)^{3}}\cos^{2}\theta+\mathellipsis \;,
\label{em1}
\end{equation}
where $\theta$ is the angle between $\mathbf{p}$ and $\hbar\mathbf{k}$. After taking the sum over $\mathbf{k}$, we identify the first term in Eq.~(\ref{em1}) as the second order correction to the polaron energy $\mu$ while the second term vanishes due to symmetry. Thus, one ends up with
\begin{equation*}
E_{0}^{(2)}-\mu+gn=\frac{-g^{2}n}{L^{2}}\sum_{\mathbf{k}}\left[\frac{\left(\mathbf{k}\xi\right)^{2}}{\left(\mathbf{k}\xi\right)^{2}+2}\right]^{1/2}\frac{\hbar^{2}k^{2}\cos^{2}\theta p^{2}/m^2}{\left(\frac{\hbar^{2}\theta\mathbf{k^{2}}}{2m}+\epsilon_{\mathbf{k}}\right)^{3}}\;,
\end{equation*}
and by performing the integration over momenta one finds
\begin{equation*}
E_{0}^{(2)}-\mu+gn=-\frac{1}{2}\frac{\ln\left(1/na_{B}^{2}\right)}{\ln^2(1/na^{2})}\frac{p^{2}}{2m}\;.
\end{equation*}
The result (\ref{effectivemassdef}) is then given by
\begin{equation}\tag{A6}
\frac{m}{m^\ast}=1-\frac{1}{2}\frac{\ln\left(1/na_{B}^{2}\right)}{\ln^2(1/na^{2})}\;.\\
\end{equation}
Finally, in terms of  the wavevector $k_{F}$, the ratio $m/m^\ast$ is written as in Eq.[6] of the main text.\\

\textit{Quasiparticle residue:} Within perturbation theory one computes the correction to the ground state as
\begin{equation}\tag{A7}
\left|\mathbf{p},0\rangle{}_{\text{pert}}\right.=\left|0\right\rangle +\sum_{\mathbf{k}\neq0}\frac{\left\langle\mathbf{k}\right|H_{\text{int}}\left|0\right\rangle}{E_{0}^{(0)}-E_{\mathbf{k}}^{(0)}}\left|\mathbf{k}\right\rangle +\mathellipsis,
\end{equation}
where we neglect terms of second order in $H_{\text{int}}$ orthogonal to $|0\rangle$, as well as higher order contributions. The quasiparticle residue is defined as the square modulus of the overlap between the unperturbed state $\left|0\right\rangle$  and the normalized perturbed state $\sqrt{Z_{\mathbf{p}}}\left|\mathbf{p},0\right\rangle_{\text{pert}}$. Up to second order contributions in $H_{\text{int}}$ one finds
\begin{equation*}
Z_{\mathbf{p}}=\frac{1}{{}_{\text{pert}}^{}\left\langle\mathbf{p},0\right.\left|\mathbf{p},0\right\rangle{}_{\text{pert}}^{}}=1-\sum_{\mathbf{k}\neq0}\frac{\left|\left\langle\mathbf{k}\right|H_{\text{int}}\left|0\right\rangle \right|^{2}}{\left(E_{0}^{(0)}-E_{\mathbf{k}}^{(0)}\right)^{2}}\;.
\end{equation*}
By carrying out the integral over momenta and by taking the limit $\mathbf{p}\to0$ similarly to the case of the effective mass, one finds
\begin{equation}\tag{A8}
Z_0=1-\frac{\ln\left(1/na_{B}^{2}\right)}{\ln^{2}\left(1/na^{2}\right)}\;.
\end{equation}
The above result, if written in terms of  the wavevector $k_{F}$, reduces to Eq.~[7] of the main text.\\

\section{APPENDIX B: TRIAL WAVE FUNCTIONS AND INTER ATOMIC POTENTIALS}

We describe the trial wave function which is used in QMC simulations as a guiding function for importance sampling and to impose proper boundary conditions on the many-body state. In general, the trial wave function is written as a pair product of Jastrow functions
\begin{equation}\tag{B1}
\psi_T({\bf R})=\prod_{i<j}f_B(r_{ij})\prod_{i=1}^N f_I(r_{i\alpha})\;,
\label{trial1}
\end{equation}
where ${\bf R}=({\bf r}_\alpha,{\bf r}_1,\dots,{\bf r}_N)$ is the multidimensional vector containing the spatial coordinates of the impurity and of the bath particles and $f_B$ and $f_I$ are two-body terms accounting, respectively, for boson-boson and impurity-boson correlations. 

As a general strategy, the short-range part of both the boson-boson and boson-impurity Jastrow function is taken from the lowest energy solution of the two-body scattering problem $-\frac{\hbar^{2}}{2m_{r}}\nabla^{2}\psi({\bf r})+V(r)\psi({\bf r})=0$, where $V(r)$ is the corresponding interaction potential and $m_r$ is the reduced mass. Notice that the impurity is considered to have the same mass as the bath particles yielding in both cases $2m_{r}=m$. The two-body short-range behavior is matched with an appropriate tail at large distances specific for boson-boson and boson-impurity correlations. 

\subsection{Boson-boson Jastrow terms}

Boson-boson interactions are modelled via a repulsive soft-disk potential $V_{B}(r)=V_{0}\Theta(R_{0}-r)$ of diameter $R_0$, where $\Theta(x)$ is the Heaviside function. The scattering length $a_{B}$ is related to the range $R_0$ and the height $V_0>0$ of the potential according to: $a_{B}=R_{0}\exp\left[-\frac{1}{k_{0}R_{0}}\frac{I_{0}(k_{0}R_{0})}{I_{1}(k_{0}R_{0})}\right]$. Here, $k_{0}=\sqrt{V_{0}m/\hbar^{2}}$ is the characteristic momentum associated to the potential and $I_{l}$ is the modified Bessel function of zeroth $(l=0)$ and first order $(l=1)$. In our calculations we use $nR_0^2 = 0.01$, thus ensuring that $R_0$ is small compared to the mean interparticle distance. The value of the 2D scattering length $a_B$ is exponentially suppressed and allows us to describe typical experimental conditions where the 3D $s$-wave scattering length is much smaller than the transverse length of the 2D confinement~\cite{Bloch08}. In particular, we choose the height $V_0$ of the repulsive potential such that the 2D gas parameter is equal to $na_B^2=10^{-40}$. This corresponds to a dimensionless coupling constant of the bath $\tilde{g}_B=\frac{mg_B}{\hbar^2}\simeq0.136$, quite close to the experimental conditions of Ref.~\cite{Ville18}.

The Jastrow term for boson-boson correlations is chosen of the following form,
\begin{equation}\tag{B2}
f_B(r)=\left\{ \begin{array}{c}
I_{0}(k_{0}r)\qquad r<R_{0}\\
\\
A\ln\left(\frac{r}{a_{B}}\right)\qquad R_{0}\leq r<\overline{R}\\
\\
B\exp\left(-\frac{C}{r}+\frac{D}{r^{2}}\right)\quad\overline{R}\leq r<L/2 \;.
\end{array}\right.
\end{equation}
Here, $A=I_0(k_0R_0)/\ln(R_0/a_B)$ to ensure continuity of $f_B(r)$ at $r=R_0$. Furthermore, the coefficients $B$, $C$ and $D$ are chosen such that $f_B$ and its first derivative $f_B^\prime$ are continuous functions at the matching point $\overline{R}$ and $f_B^\prime(r=L/2)=0$, complying with the periodic boundary conditions. The position $\overline{R}$ of the matching point is a parameter optimized by minimizing the energy in a variational calculation. As stated, the short-range part corresponds to the two-body scattering solution at zero energy and the leading long-range part reproduces the phononic tail as predicted from hydrodynamic theory~\cite{ReattoChester67}. 

\subsection{Impurity-boson Jastrow terms: Attractive and repulsive branch}

The impurity-boson interaction is modelled by a contact pseudo-potential. In this case the interaction potential is replaced by Bethe-Peierls boundary conditions on the many-body wave function when a particle of the bath approaches the impurity. These contact conditions are imposed by the term $\prod_{i=1}^N f_I(r_{i\alpha})$ in the trial function~(\ref{trial1}). Notice that the pseudo-potential supports a two-body bound state for any value of the scattering length $a$.  The energy of this bound state is given by $\epsilon_{b}=-\frac{4e^{-2\gamma}\hbar^2}{ma^2}$, where $\gamma=0.577$ is Euler's constant.

For the attractive branch the correlation term $f_I(r)$ is constructed in the following way
\begin{equation}\tag{B3}
f_I(r)=\left\{ \begin{array}{cc}
AK_{0}(2e^{-\gamma}r/a),&r\leq\overline{R_1}\\
\\
B + e^{-Cr}
+e^{-C(L-r)},&\overline{R_1}\leq r<L/2\;.
\end{array}\right.
\label{Eq:wfABattractive}
\end{equation}
Here, $K_0$ is the modified Bessel function of the second kind. The parameters $A$ and $B$ are chosen such that $f_I(r)$ and its first derivative are continuous at $r=\overline{R_1}$. The parameter $C$ and the matching point $\overline{R_1}$ are instead additional parameters used to minimize the variational energy. Notice that, by construction, $f_I^\prime(r=L/2)=0$ in compliance with periodic boundary conditions.

The Jastrow term describing the repulsive branch is instead chosen as
\begin{equation}\tag{B4}
f_I(r)=\left\{ \begin{array}{c}
A\ln\left(\frac{r}{a}\right), \qquad r \leq\overline{R_1}\\
\\
B + e^{-Cr}
+e^{-C(L-r)}, \quad\overline{R_1}\leq r<L/2 \;.
\end{array}\right.
\label{Eq:wfABrepulsive}
\end{equation}
The parameters $A$, $B$, $C$ and $\overline{R_1}$ entering Eq.~(\ref{Eq:wfABrepulsive}) are chosen similarly to Eq.~(\ref{Eq:wfABattractive}).

We point out that, for both branches, the pair wave function satisfies the 2D Bethe-Peierls contact condition of a pseudo-potential with scattering length $a$ which reads: $\frac{[rf_I^\prime]_{r=0}}{[f_I-\ln(qr)rf_I^\prime]_{r=0}}=-\frac{1}{\ln(qa)}$, where $q$ is an arbitrary wave vector. The important difference between attractive and repulsive branches is that in the former case the function $f_I$ is nodeless and properly describes the ground state of the polaron. In the latter case, instead, $f_I$ has a node at $r=a$ and its short-distance behavior corresponds to an excited state of the two-body problem orthogonal to the bound state with energy $\epsilon_b$. 

We also notice that the long-range behavior of the Jastrow term $f_I(r)$ is consistent with perturbation theory applied to the Gross-Pitaevskii equation of a Bose condensate in presence of a quenched impurity with infinite mass. In fact, it can be shown~\cite{Astrakharchik04,AstrakharchikPhD} that such impurity induces the following perturbation, $\delta\psi_{\bf k}$, to the wave function of the bath in momentum space,
\begin{equation}\tag{B5}
\delta\psi_{{\bf k}} = -\frac{g\phi_{0}}{\frac{\hbar ^{2}k^{2}}{2m}+2mc^{2}}\, , \label{dPsi}
\end{equation}
where $c=\sqrt{g_Bn/m}$ is the speed of sound in the bath and $\phi_0$ is the wave function of the unperturbed condensate.
In coordinate space, the perturbation decays exponentially in 3D and 1D with the healing length given by $\xi=\hbar/(\sqrt{2}mc)$, while in 2D it involves the modified Bessel function of the second kind
\begin{equation}\tag{B6}
\delta \psi(r) \propto \frac{mg\phi_0}{\hbar^2} K_0\left(\frac{\sqrt{2}r}{\xi}\right)\;.
\end{equation}
The above expression can be expanded with logarithmic accuracy at large distances as
\begin{equation}\tag{B7}
\delta \psi(r) \propto \frac{mg\phi_0}{\hbar^2} \exp\left[-\frac{\sqrt{2}r}{\xi} + {\cal O}\left(\ln\frac{\xi}{r}\right)\right]\;,
\label{Eq:df:polaron}
\end{equation}
exhibiting the same functional form as the long-range behavior in Eqs.~(\ref{Eq:wfABattractive}) and (\ref{Eq:wfABrepulsive}).

In addition we have also used a square-well potential with fixed (short) radius to model the boson-impurity interaction. We find that the obtained results depend only on the interaction strength $\ln(k_Fa)$ and not on the details of the potential.

\section{APPENDIX C: REPULSIVE POLARON BRANCH}

\begin{figure}
\begin{center}
\includegraphics[width=1\linewidth]{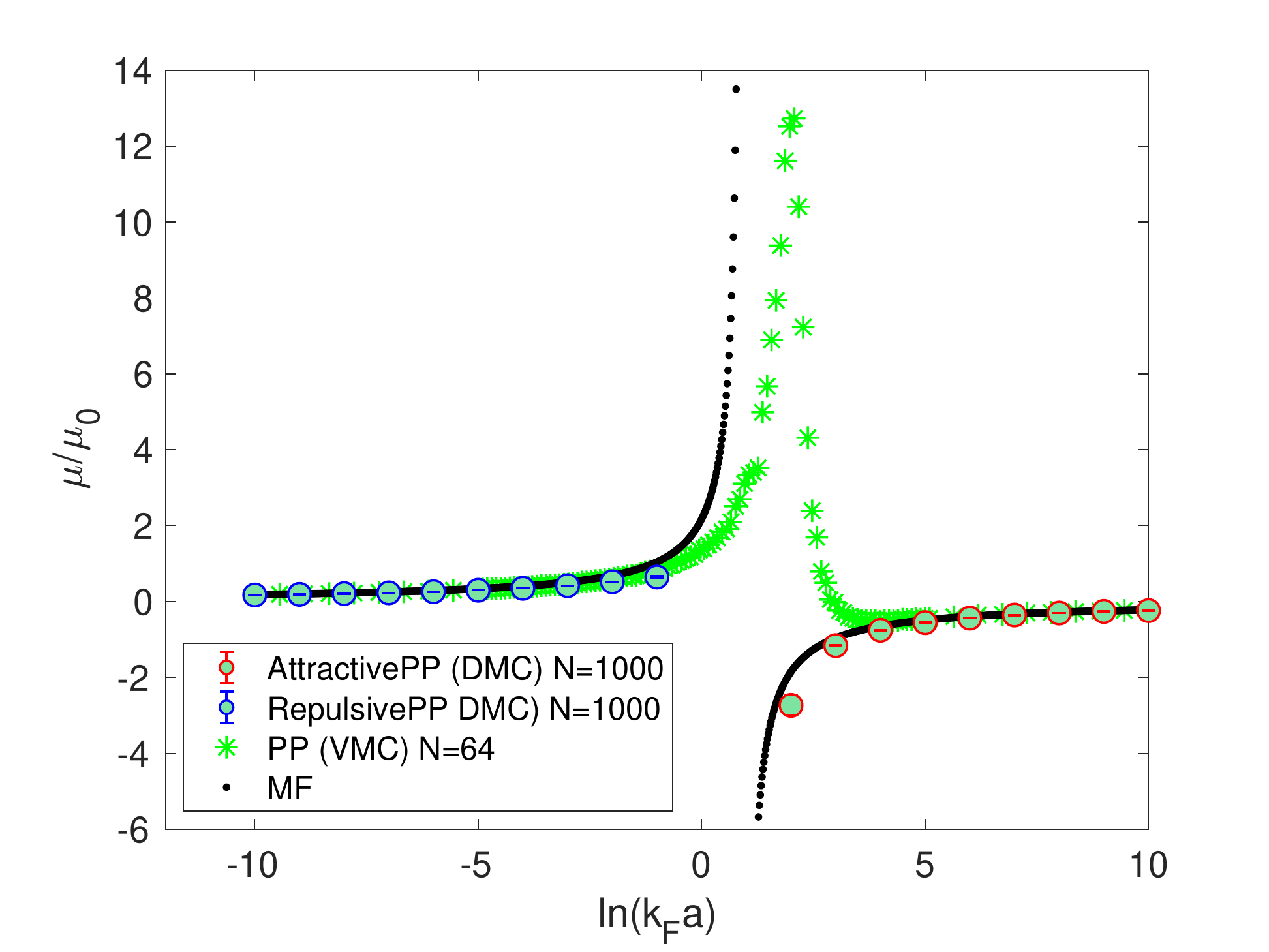}
\caption{Variational Monte Carlo (VMC) results for the polaron energy interpolating between the attractive and repulsive branch. Circles show the Diffusion Monte Carlo (DMC) results reported in Fig.~2 in the main text. Solid lines correspond to the perturbation theory results in Eq.~(5) of the main text.}
\label{Fig:polrep}
\end{center}
\end{figure}

An additional physical insight can be obtained from variational calculations for the excited branch using the Jastrow term in Eq.~(\ref{Eq:wfABrepulsive}). Note that this Jastrow term has a node when $r=a$ and if $a\ll\overline{R_1}$ it is orthogonal to the bound state. This choice of trial many-body wave function describes an excited state of the polaron which is expected to be metastable when the mean interparticle distance is much larger than $a$, i.e. for $|\ln(k_Fa)|\ll 1$.
In the opposite limit when the relevant distances are much smaller than $a$, the attractive~(\ref{Eq:wfABattractive}) and repulsive~(\ref{Eq:wfABrepulsive}) Jastrow terms give the same function, in fact $K_0(2e^{-\gamma}r/a) \approx \ln(r/a)$ if $r\ll a$. This means that, at the variational level, the upper repulsive branch constructed from the Jastrow term~(\ref{Eq:wfABrepulsive}) connects with the lower attractive branch~(\ref{Eq:wfABattractive}) for $|\ln(k_Fa)|\gg 1$, see Fig.~\ref{Fig:polrep}. Notice, however, that these variational estimates are upper bounds to the true ground-state energy, which is large and negative and corresponds to a deep bound state of the impurity and many particles from the bath.

In other words, the pair-product construction in Eq.~(\ref{Eq:wfABrepulsive}) for the upper branch becomes unstable due to a significant overlap with the bound state. Although the exact position of the crossing from the upper to the lower branch is not expected to be quantitatively correct, its presence hints to a possible instability of the upper branch also in experimentally relevant configurations with ultracold atoms.

\bibliography{BOSEPOL2D}

\begin{thebibliography}{71}%
\makeatletter
\providecommand \@ifxundefined [1]{%
 \@ifx{#1\undefined}
}%
\providecommand \@ifnum [1]{%
 \ifnum #1\expandafter \@firstoftwo
 \else \expandafter \@secondoftwo
 \fi
}%
\providecommand \@ifx [1]{%
 \ifx #1\expandafter \@firstoftwo
 \else \expandafter \@secondoftwo
 \fi
}%
\providecommand \natexlab [1]{#1}%
\providecommand \enquote  [1]{``#1''}%
\providecommand \bibnamefont  [1]{#1}%
\providecommand \bibfnamefont [1]{#1}%
\providecommand \citenamefont [1]{#1}%
\providecommand \href@noop [0]{\@secondoftwo}%
\providecommand \href [0]{\begingroup \@sanitize@url \@href}%
\providecommand \@href[1]{\@@startlink{#1}\@@href}%
\providecommand \@@href[1]{\endgroup#1\@@endlink}%
\providecommand \@sanitize@url [0]{\catcode `\\12\catcode `\$12\catcode
  `\&12\catcode `\#12\catcode `\^12\catcode `\_12\catcode `\%12\relax}%
\providecommand \@@startlink[1]{}%
\providecommand \@@endlink[0]{}%
\providecommand \url  [0]{\begingroup\@sanitize@url \@url }%
\providecommand \@url [1]{\endgroup\@href {#1}{\urlprefix }}%
\providecommand \urlprefix  [0]{URL }%
\providecommand \Eprint [0]{\href }%
\providecommand \doibase [0]{http://dx.doi.org/}%
\providecommand \selectlanguage [0]{\@gobble}%
\providecommand \bibinfo  [0]{\@secondoftwo}%
\providecommand \bibfield  [0]{\@secondoftwo}%
\providecommand \translation [1]{[#1]}%
\providecommand \BibitemOpen [0]{}%
\providecommand \bibitemStop [0]{}%
\providecommand \bibitemNoStop [0]{.\EOS\space}%
\providecommand \EOS [0]{\spacefactor3000\relax}%
\providecommand \BibitemShut  [1]{\csname bibitem#1\endcsname}%
\let\auto@bib@innerbib\@empty
\bibitem [{\citenamefont {Landau}\ and\ \citenamefont
  {Pekar}(1948)}]{Landau48}%
  \BibitemOpen
  \bibfield  {author} {\bibinfo {author} {\bibfnamefont {L.}~\bibnamefont
  {Landau}}\ and\ \bibinfo {author} {\bibfnamefont {S.}~\bibnamefont {Pekar}},\
  }\href@noop {} {\bibfield  {journal} {\bibinfo  {journal} {J. Exp. Theor.
  Phys}\ }\textbf {\bibinfo {volume} {18}},\ \bibinfo {pages} {419} (\bibinfo
  {year} {1948})}\BibitemShut {NoStop}%
\bibitem [{\citenamefont {Devreese}\ and\ \citenamefont
  {Peters}(1984)}]{PolaronsandExcitonsBook}%
  \BibitemOpen
  \bibfield  {author} {\bibinfo {author} {\bibfnamefont {J.}~\bibnamefont
  {Devreese}}\ and\ \bibinfo {author} {\bibfnamefont {F.}~\bibnamefont
  {Peters}},\ }\href@noop {} {\emph {\bibinfo {title} {Polarons and Excitons in
  Polar Semiconductors and Ionic Crystals}}}\ (\bibinfo  {publisher} {Plenum
  Press, New York},\ \bibinfo {year} {1984})\BibitemShut {NoStop}%
\bibitem [{\citenamefont {Gershenson}\ \emph {et~al.}(2006)\citenamefont
  {Gershenson}, \citenamefont {Podzorov},\ and\ \citenamefont
  {Morpurgo}}]{Gershenson78}%
  \BibitemOpen
  \bibfield  {author} {\bibinfo {author} {\bibfnamefont {M.~E.}\ \bibnamefont
  {Gershenson}}, \bibinfo {author} {\bibfnamefont {V.}~\bibnamefont
  {Podzorov}}, \ and\ \bibinfo {author} {\bibfnamefont {A.~F.}\ \bibnamefont
  {Morpurgo}},\ }\href {\doibase 10.1103/RevModPhys.78.973} {\bibfield
  {journal} {\bibinfo  {journal} {Rev. Mod. Phys.}\ }\textbf {\bibinfo {volume}
  {78}},\ \bibinfo {pages} {973} (\bibinfo {year} {2006})}\BibitemShut
  {NoStop}%
\bibitem [{\citenamefont {Watanabe}\ \emph {et~al.}(2014)\citenamefont
  {Watanabe}, \citenamefont {Ando}, \citenamefont {Kang}, \citenamefont
  {Mooser}, \citenamefont {Vaynzof}, \citenamefont {Kurebayashi}, \citenamefont
  {Saitoh},\ and\ \citenamefont {Sirringhaus}}]{Watanabe14}%
  \BibitemOpen
  \bibfield  {author} {\bibinfo {author} {\bibfnamefont {S.}~\bibnamefont
  {Watanabe}}, \bibinfo {author} {\bibfnamefont {K.}~\bibnamefont {Ando}},
  \bibinfo {author} {\bibfnamefont {K.}~\bibnamefont {Kang}}, \bibinfo {author}
  {\bibfnamefont {S.}~\bibnamefont {Mooser}}, \bibinfo {author} {\bibfnamefont
  {Y.}~\bibnamefont {Vaynzof}}, \bibinfo {author} {\bibfnamefont
  {H.}~\bibnamefont {Kurebayashi}}, \bibinfo {author} {\bibfnamefont
  {E.}~\bibnamefont {Saitoh}}, \ and\ \bibinfo {author} {\bibfnamefont
  {H.}~\bibnamefont {Sirringhaus}},\ }\href@noop {} {\bibfield  {journal}
  {\bibinfo  {journal} {Nat. Phys.}\ }\textbf {\bibinfo {volume} {10}},\
  \bibinfo {pages} {308} (\bibinfo {year} {2014})}\BibitemShut {NoStop}%
\bibitem [{\citenamefont {Lee}\ \emph {et~al.}(2006)\citenamefont {Lee},
  \citenamefont {Nagaosa},\ and\ \citenamefont {Wen}}]{Lee06}%
  \BibitemOpen
  \bibfield  {author} {\bibinfo {author} {\bibfnamefont {P.~A.}\ \bibnamefont
  {Lee}}, \bibinfo {author} {\bibfnamefont {N.}~\bibnamefont {Nagaosa}}, \ and\
  \bibinfo {author} {\bibfnamefont {X.-G.}\ \bibnamefont {Wen}},\ }\href
  {\doibase 10.1103/RevModPhys.78.17} {\bibfield  {journal} {\bibinfo
  {journal} {Rev. Mod. Phys.}\ }\textbf {\bibinfo {volume} {78}},\ \bibinfo
  {pages} {17} (\bibinfo {year} {2006})}\BibitemShut {NoStop}%
\bibitem [{\citenamefont {Gutierrez}\ and\ \citenamefont
  {Cuniberti}(2008)}]{EffectiveModelsforChargeTransportinDNANanowires}%
  \BibitemOpen
  \bibfield  {author} {\bibinfo {author} {\bibfnamefont {R.}~\bibnamefont
  {Gutierrez}}\ and\ \bibinfo {author} {\bibfnamefont {G.}~\bibnamefont
  {Cuniberti}},\ }\href@noop {} {\bibfield  {journal} {\bibinfo  {journal}
  {NanoBioTechnology}\ } (\bibinfo {year} {2008})}\BibitemShut {NoStop}%
\bibitem [{\citenamefont {Baym}\ and\ \citenamefont
  {Pethick}(1991)}]{LandauFermiLiquidTheory}%
  \BibitemOpen
  \bibfield  {author} {\bibinfo {author} {\bibfnamefont {G.}~\bibnamefont
  {Baym}}\ and\ \bibinfo {author} {\bibfnamefont {C.}~\bibnamefont {Pethick}},\
  }\href@noop {} {\emph {\bibinfo {title} {Landau Fermi-Liquid Theory: Concepts
  and Applications}}}\ (\bibinfo  {publisher} {Wiley-VCH-New York.},\ \bibinfo
  {year} {1991})\BibitemShut {NoStop}%
\bibitem [{\citenamefont {Bloch}\ \emph {et~al.}(2008)\citenamefont {Bloch},
  \citenamefont {Dalibard},\ and\ \citenamefont {Zwerger}}]{Bloch08}%
  \BibitemOpen
  \bibfield  {author} {\bibinfo {author} {\bibfnamefont {I.}~\bibnamefont
  {Bloch}}, \bibinfo {author} {\bibfnamefont {J.}~\bibnamefont {Dalibard}}, \
  and\ \bibinfo {author} {\bibfnamefont {W.}~\bibnamefont {Zwerger}},\ }\href
  {\doibase 10.1103/RevModPhys.80.885} {\bibfield  {journal} {\bibinfo
  {journal} {Rev. Mod. Phys.}\ }\textbf {\bibinfo {volume} {80}},\ \bibinfo
  {pages} {885} (\bibinfo {year} {2008})}\BibitemShut {NoStop}%
\bibitem [{\citenamefont {Schirotzek}\ \emph {et~al.}(2009)\citenamefont
  {Schirotzek}, \citenamefont {Wu}, \citenamefont {Sommer},\ and\ \citenamefont
  {Zwierlein}}]{Schirotzek09}%
  \BibitemOpen
  \bibfield  {author} {\bibinfo {author} {\bibfnamefont {A.}~\bibnamefont
  {Schirotzek}}, \bibinfo {author} {\bibfnamefont {C.-H.}\ \bibnamefont {Wu}},
  \bibinfo {author} {\bibfnamefont {A.}~\bibnamefont {Sommer}}, \ and\ \bibinfo
  {author} {\bibfnamefont {M.~W.}\ \bibnamefont {Zwierlein}},\ }\href {\doibase
  10.1103/PhysRevLett.102.230402} {\bibfield  {journal} {\bibinfo  {journal}
  {Phys. Rev. Lett.}\ }\textbf {\bibinfo {volume} {102}},\ \bibinfo {pages}
  {230402} (\bibinfo {year} {2009})}\BibitemShut {NoStop}%
\bibitem [{\citenamefont {Koschorreck}\ \emph {et~al.}(2012)\citenamefont
  {Koschorreck}, \citenamefont {Pertot}, \citenamefont {Vogt}, \citenamefont
  {Fr{\"o}hlich}, \citenamefont {Feld},\ and\ \citenamefont
  {K{\"o}hl}}]{Koschorreck12}%
  \BibitemOpen
  \bibfield  {author} {\bibinfo {author} {\bibfnamefont {M.}~\bibnamefont
  {Koschorreck}}, \bibinfo {author} {\bibfnamefont {D.}~\bibnamefont {Pertot}},
  \bibinfo {author} {\bibfnamefont {E.}~\bibnamefont {Vogt}}, \bibinfo {author}
  {\bibfnamefont {B.}~\bibnamefont {Fr{\"o}hlich}}, \bibinfo {author}
  {\bibfnamefont {M.}~\bibnamefont {Feld}}, \ and\ \bibinfo {author}
  {\bibfnamefont {M.}~\bibnamefont {K{\"o}hl}},\ }\href@noop {} {\bibfield
  {journal} {\bibinfo  {journal} {Nature}\ }\textbf {\bibinfo {volume} {485}},\
  \bibinfo {pages} {619} (\bibinfo {year} {2012})}\BibitemShut {NoStop}%
\bibitem [{\citenamefont {Kohstall}\ \emph {et~al.}(2012)\citenamefont
  {Kohstall}, \citenamefont {Zaccanti}, \citenamefont {Jag}, \citenamefont
  {Trenkwalder}, \citenamefont {Massignan}, \citenamefont {Bruun},
  \citenamefont {Schreck},\ and\ \citenamefont {Grimm}}]{Kohstall12}%
  \BibitemOpen
  \bibfield  {author} {\bibinfo {author} {\bibfnamefont {C.}~\bibnamefont
  {Kohstall}}, \bibinfo {author} {\bibfnamefont {M.}~\bibnamefont {Zaccanti}},
  \bibinfo {author} {\bibfnamefont {M.}~\bibnamefont {Jag}}, \bibinfo {author}
  {\bibfnamefont {A.}~\bibnamefont {Trenkwalder}}, \bibinfo {author}
  {\bibfnamefont {P.}~\bibnamefont {Massignan}}, \bibinfo {author}
  {\bibfnamefont {G.~M.}\ \bibnamefont {Bruun}}, \bibinfo {author}
  {\bibfnamefont {F.}~\bibnamefont {Schreck}}, \ and\ \bibinfo {author}
  {\bibfnamefont {R.}~\bibnamefont {Grimm}},\ }\href@noop {} {\bibfield
  {journal} {\bibinfo  {journal} {Nature}\ }\textbf {\bibinfo {volume} {485}},\
  \bibinfo {pages} {615} (\bibinfo {year} {2012})}\BibitemShut {NoStop}%
\bibitem [{\citenamefont {Cetina}\ \emph {et~al.}(2016)\citenamefont {Cetina},
  \citenamefont {Jag}, \citenamefont {Lous}, \citenamefont {Fritsche},
  \citenamefont {Walraven}, \citenamefont {Grimm}, \citenamefont {Levinsen},
  \citenamefont {Parish}, \citenamefont {Schmidt}, \citenamefont {Knap},\ and\
  \citenamefont {Demler}}]{Cetina16}%
  \BibitemOpen
  \bibfield  {author} {\bibinfo {author} {\bibfnamefont {M.}~\bibnamefont
  {Cetina}}, \bibinfo {author} {\bibfnamefont {M.}~\bibnamefont {Jag}},
  \bibinfo {author} {\bibfnamefont {R.~S.}\ \bibnamefont {Lous}}, \bibinfo
  {author} {\bibfnamefont {I.}~\bibnamefont {Fritsche}}, \bibinfo {author}
  {\bibfnamefont {J.~T.}\ \bibnamefont {Walraven}}, \bibinfo {author}
  {\bibfnamefont {R.}~\bibnamefont {Grimm}}, \bibinfo {author} {\bibfnamefont
  {J.}~\bibnamefont {Levinsen}}, \bibinfo {author} {\bibfnamefont {M.~M.}\
  \bibnamefont {Parish}}, \bibinfo {author} {\bibfnamefont {R.}~\bibnamefont
  {Schmidt}}, \bibinfo {author} {\bibfnamefont {M.}~\bibnamefont {Knap}}, \
  and\ \bibinfo {author} {\bibfnamefont {E.}~\bibnamefont {Demler}},\
  }\href@noop {} {\bibfield  {journal} {\bibinfo  {journal} {Science}\ }\textbf
  {\bibinfo {volume} {354}},\ \bibinfo {pages} {96} (\bibinfo {year}
  {2016})}\BibitemShut {NoStop}%
\bibitem [{\citenamefont {Scazza}\ \emph {et~al.}(2017)\citenamefont {Scazza},
  \citenamefont {Valtolina}, \citenamefont {Massignan}, \citenamefont {Recati},
  \citenamefont {Amico}, \citenamefont {Burchianti}, \citenamefont {Fort},
  \citenamefont {Inguscio}, \citenamefont {Zaccanti},\ and\ \citenamefont
  {Roati}}]{Scazza17}%
  \BibitemOpen
  \bibfield  {author} {\bibinfo {author} {\bibfnamefont {F.}~\bibnamefont
  {Scazza}}, \bibinfo {author} {\bibfnamefont {G.}~\bibnamefont {Valtolina}},
  \bibinfo {author} {\bibfnamefont {P.}~\bibnamefont {Massignan}}, \bibinfo
  {author} {\bibfnamefont {A.}~\bibnamefont {Recati}}, \bibinfo {author}
  {\bibfnamefont {A.}~\bibnamefont {Amico}}, \bibinfo {author} {\bibfnamefont
  {A.}~\bibnamefont {Burchianti}}, \bibinfo {author} {\bibfnamefont
  {C.}~\bibnamefont {Fort}}, \bibinfo {author} {\bibfnamefont {M.}~\bibnamefont
  {Inguscio}}, \bibinfo {author} {\bibfnamefont {M.}~\bibnamefont {Zaccanti}},
  \ and\ \bibinfo {author} {\bibfnamefont {G.}~\bibnamefont {Roati}},\ }\href
  {\doibase 10.1103/PhysRevLett.118.083602} {\bibfield  {journal} {\bibinfo
  {journal} {Phys. Rev. Lett.}\ }\textbf {\bibinfo {volume} {118}},\ \bibinfo
  {pages} {083602} (\bibinfo {year} {2017})}\BibitemShut {NoStop}%
\bibitem [{\citenamefont {J\o{}rgensen}\ \emph {et~al.}(2016)\citenamefont
  {J\o{}rgensen}, \citenamefont {Wacker}, \citenamefont {Skalmstang},
  \citenamefont {Parish}, \citenamefont {Levinsen}, \citenamefont
  {Christensen}, \citenamefont {Bruun},\ and\ \citenamefont
  {Arlt}}]{Jorgensen16}%
  \BibitemOpen
  \bibfield  {author} {\bibinfo {author} {\bibfnamefont {N.~B.}\ \bibnamefont
  {J\o{}rgensen}}, \bibinfo {author} {\bibfnamefont {L.}~\bibnamefont
  {Wacker}}, \bibinfo {author} {\bibfnamefont {K.~T.}\ \bibnamefont
  {Skalmstang}}, \bibinfo {author} {\bibfnamefont {M.~M.}\ \bibnamefont
  {Parish}}, \bibinfo {author} {\bibfnamefont {J.}~\bibnamefont {Levinsen}},
  \bibinfo {author} {\bibfnamefont {R.~S.}\ \bibnamefont {Christensen}},
  \bibinfo {author} {\bibfnamefont {G.~M.}\ \bibnamefont {Bruun}}, \ and\
  \bibinfo {author} {\bibfnamefont {J.~J.}\ \bibnamefont {Arlt}},\ }\href
  {\doibase 10.1103/PhysRevLett.117.055302} {\bibfield  {journal} {\bibinfo
  {journal} {Phys. Rev. Lett.}\ }\textbf {\bibinfo {volume} {117}},\ \bibinfo
  {pages} {055302} (\bibinfo {year} {2016})}\BibitemShut {NoStop}%
\bibitem [{\citenamefont {Hu}\ \emph {et~al.}(2016)\citenamefont {Hu},
  \citenamefont {Van~de Graaff}, \citenamefont {Kedar}, \citenamefont {Corson},
  \citenamefont {Cornell},\ and\ \citenamefont {Jin}}]{Hu16}%
  \BibitemOpen
  \bibfield  {author} {\bibinfo {author} {\bibfnamefont {M.-G.}\ \bibnamefont
  {Hu}}, \bibinfo {author} {\bibfnamefont {M.~J.}\ \bibnamefont {Van~de
  Graaff}}, \bibinfo {author} {\bibfnamefont {D.}~\bibnamefont {Kedar}},
  \bibinfo {author} {\bibfnamefont {J.~P.}\ \bibnamefont {Corson}}, \bibinfo
  {author} {\bibfnamefont {E.~A.}\ \bibnamefont {Cornell}}, \ and\ \bibinfo
  {author} {\bibfnamefont {D.~S.}\ \bibnamefont {Jin}},\ }\href {\doibase
  10.1103/PhysRevLett.117.055301} {\bibfield  {journal} {\bibinfo  {journal}
  {Phys. Rev. Lett.}\ }\textbf {\bibinfo {volume} {117}},\ \bibinfo {pages}
  {055301} (\bibinfo {year} {2016})}\BibitemShut {NoStop}%
\bibitem [{\citenamefont {Camargo}\ \emph {et~al.}(2018)\citenamefont
  {Camargo}, \citenamefont {Schmidt}, \citenamefont {Whalen}, \citenamefont
  {Ding}, \citenamefont {Woehl}, \citenamefont {Yoshida}, \citenamefont
  {Burgd\"orfer}, \citenamefont {Dunning}, \citenamefont {Sadeghpour},
  \citenamefont {Demler},\ and\ \citenamefont {Killian}}]{Camargo18}%
  \BibitemOpen
  \bibfield  {author} {\bibinfo {author} {\bibfnamefont {F.}~\bibnamefont
  {Camargo}}, \bibinfo {author} {\bibfnamefont {R.}~\bibnamefont {Schmidt}},
  \bibinfo {author} {\bibfnamefont {J.~D.}\ \bibnamefont {Whalen}}, \bibinfo
  {author} {\bibfnamefont {R.}~\bibnamefont {Ding}}, \bibinfo {author}
  {\bibfnamefont {G.}~\bibnamefont {Woehl}}, \bibinfo {author} {\bibfnamefont
  {S.}~\bibnamefont {Yoshida}}, \bibinfo {author} {\bibfnamefont
  {J.}~\bibnamefont {Burgd\"orfer}}, \bibinfo {author} {\bibfnamefont {F.~B.}\
  \bibnamefont {Dunning}}, \bibinfo {author} {\bibfnamefont {H.~R.}\
  \bibnamefont {Sadeghpour}}, \bibinfo {author} {\bibfnamefont
  {E.}~\bibnamefont {Demler}}, \ and\ \bibinfo {author} {\bibfnamefont {T.~C.}\
  \bibnamefont {Killian}},\ }\href {\doibase 10.1103/PhysRevLett.120.083401}
  {\bibfield  {journal} {\bibinfo  {journal} {Phys. Rev. Lett.}\ }\textbf
  {\bibinfo {volume} {120}},\ \bibinfo {pages} {083401} (\bibinfo {year}
  {2018})}\BibitemShut {NoStop}%
\bibitem [{\citenamefont {Pe\~na Ardila}\ \emph {et~al.}(2019)\citenamefont
  {Pe\~na Ardila}, \citenamefont {J\o{}rgensen}, \citenamefont {Pohl},
  \citenamefont {Giorgini}, \citenamefont {Bruun},\ and\ \citenamefont
  {Arlt}}]{Ardila18}%
  \BibitemOpen
  \bibfield  {author} {\bibinfo {author} {\bibfnamefont {L.~A.}\ \bibnamefont
  {Pe\~na Ardila}}, \bibinfo {author} {\bibfnamefont {N.~B.}\ \bibnamefont
  {J\o{}rgensen}}, \bibinfo {author} {\bibfnamefont {T.}~\bibnamefont {Pohl}},
  \bibinfo {author} {\bibfnamefont {S.}~\bibnamefont {Giorgini}}, \bibinfo
  {author} {\bibfnamefont {G.~M.}\ \bibnamefont {Bruun}}, \ and\ \bibinfo
  {author} {\bibfnamefont {J.~J.}\ \bibnamefont {Arlt}},\ }\href {\doibase
  10.1103/PhysRevA.99.063607} {\bibfield  {journal} {\bibinfo  {journal} {Phys.
  Rev. A}\ }\textbf {\bibinfo {volume} {99}},\ \bibinfo {pages} {063607}
  (\bibinfo {year} {2019})}\BibitemShut {NoStop}%
\bibitem [{\citenamefont {Yan}\ \emph {et~al.}(2019)\citenamefont {Yan},
  \citenamefont {Ni}, \citenamefont {Robens},\ and\ \citenamefont
  {Zwierlein}}]{Zwierlein19}%
  \BibitemOpen
  \bibfield  {author} {\bibinfo {author} {\bibfnamefont {Z.~Z.}\ \bibnamefont
  {Yan}}, \bibinfo {author} {\bibfnamefont {Y.}~\bibnamefont {Ni}}, \bibinfo
  {author} {\bibfnamefont {C.}~\bibnamefont {Robens}}, \ and\ \bibinfo {author}
  {\bibfnamefont {M.~W.}\ \bibnamefont {Zwierlein}},\ }\href@noop {} {\bibfield
   {journal} {\bibinfo  {journal} {preprint, ArXiv:1904.02685}\ } (\bibinfo
  {year} {2019})}\BibitemShut {NoStop}%
\bibitem [{\citenamefont {Spethmann}\ \emph {et~al.}(2012)\citenamefont
  {Spethmann}, \citenamefont {Kindermann}, \citenamefont {John}, \citenamefont
  {Weber}, \citenamefont {Meschede},\ and\ \citenamefont
  {Widera}}]{Spethmann12}%
  \BibitemOpen
  \bibfield  {author} {\bibinfo {author} {\bibfnamefont {N.}~\bibnamefont
  {Spethmann}}, \bibinfo {author} {\bibfnamefont {F.}~\bibnamefont
  {Kindermann}}, \bibinfo {author} {\bibfnamefont {S.}~\bibnamefont {John}},
  \bibinfo {author} {\bibfnamefont {C.}~\bibnamefont {Weber}}, \bibinfo
  {author} {\bibfnamefont {D.}~\bibnamefont {Meschede}}, \ and\ \bibinfo
  {author} {\bibfnamefont {A.}~\bibnamefont {Widera}},\ }\href {\doibase
  10.1103/PhysRevLett.109.235301} {\bibfield  {journal} {\bibinfo  {journal}
  {Phys. Rev. Lett.}\ }\textbf {\bibinfo {volume} {109}},\ \bibinfo {pages}
  {235301} (\bibinfo {year} {2012})}\BibitemShut {NoStop}%
\bibitem [{\citenamefont {Catani}\ \emph {et~al.}(2012)\citenamefont {Catani},
  \citenamefont {Lamporesi}, \citenamefont {Naik}, \citenamefont {Gring},
  \citenamefont {Inguscio}, \citenamefont {Minardi}, \citenamefont {Kantian},\
  and\ \citenamefont {Giamarchi}}]{Catani12}%
  \BibitemOpen
  \bibfield  {author} {\bibinfo {author} {\bibfnamefont {J.}~\bibnamefont
  {Catani}}, \bibinfo {author} {\bibfnamefont {G.}~\bibnamefont {Lamporesi}},
  \bibinfo {author} {\bibfnamefont {D.}~\bibnamefont {Naik}}, \bibinfo {author}
  {\bibfnamefont {M.}~\bibnamefont {Gring}}, \bibinfo {author} {\bibfnamefont
  {M.}~\bibnamefont {Inguscio}}, \bibinfo {author} {\bibfnamefont
  {F.}~\bibnamefont {Minardi}}, \bibinfo {author} {\bibfnamefont
  {A.}~\bibnamefont {Kantian}}, \ and\ \bibinfo {author} {\bibfnamefont
  {T.}~\bibnamefont {Giamarchi}},\ }\href {\doibase 10.1103/PhysRevA.85.023623}
  {\bibfield  {journal} {\bibinfo  {journal} {Phys. Rev. A}\ }\textbf {\bibinfo
  {volume} {85}},\ \bibinfo {pages} {023623} (\bibinfo {year}
  {2012})}\BibitemShut {NoStop}%
\bibitem [{\citenamefont {Viverit}\ and\ \citenamefont
  {Giorgini}(2002)}]{Viverit02}%
  \BibitemOpen
  \bibfield  {author} {\bibinfo {author} {\bibfnamefont {L.}~\bibnamefont
  {Viverit}}\ and\ \bibinfo {author} {\bibfnamefont {S.}~\bibnamefont
  {Giorgini}},\ }\href {\doibase 10.1103/PhysRevA.66.063604} {\bibfield
  {journal} {\bibinfo  {journal} {Phys. Rev. A}\ }\textbf {\bibinfo {volume}
  {66}},\ \bibinfo {pages} {063604} (\bibinfo {year} {2002})}\BibitemShut
  {NoStop}%
\bibitem [{\citenamefont {Tempere}\ \emph {et~al.}(2009)\citenamefont
  {Tempere}, \citenamefont {Casteels}, \citenamefont {Oberthaler},
  \citenamefont {Knoop}, \citenamefont {Timmermans},\ and\ \citenamefont
  {Devreese}}]{Tempere09}%
  \BibitemOpen
  \bibfield  {author} {\bibinfo {author} {\bibfnamefont {J.}~\bibnamefont
  {Tempere}}, \bibinfo {author} {\bibfnamefont {W.}~\bibnamefont {Casteels}},
  \bibinfo {author} {\bibfnamefont {M.~K.}\ \bibnamefont {Oberthaler}},
  \bibinfo {author} {\bibfnamefont {S.}~\bibnamefont {Knoop}}, \bibinfo
  {author} {\bibfnamefont {E.}~\bibnamefont {Timmermans}}, \ and\ \bibinfo
  {author} {\bibfnamefont {J.~T.}\ \bibnamefont {Devreese}},\ }\href {\doibase
  10.1103/PhysRevB.80.184504} {\bibfield  {journal} {\bibinfo  {journal} {Phys.
  Rev. B}\ }\textbf {\bibinfo {volume} {80}},\ \bibinfo {pages} {184504}
  (\bibinfo {year} {2009})}\BibitemShut {NoStop}%
\bibitem [{\citenamefont {Grusdt}\ \emph {et~al.}(2015)\citenamefont {Grusdt},
  \citenamefont {Shchadilova}, \citenamefont {Rubtsov},\ and\ \citenamefont
  {Demler}}]{Grusdt15}%
  \BibitemOpen
  \bibfield  {author} {\bibinfo {author} {\bibfnamefont {F.}~\bibnamefont
  {Grusdt}}, \bibinfo {author} {\bibfnamefont {Y.~E.}\ \bibnamefont
  {Shchadilova}}, \bibinfo {author} {\bibfnamefont {A.~N.}\ \bibnamefont
  {Rubtsov}}, \ and\ \bibinfo {author} {\bibfnamefont {E.}~\bibnamefont
  {Demler}},\ }\href@noop {} {\bibfield  {journal} {\bibinfo  {journal} {Sci.
  Rep.}\ }\textbf {\bibinfo {volume} {5}} (\bibinfo {year} {2015})}\BibitemShut
  {NoStop}%
\bibitem [{\citenamefont {{Grusdt}}\ and\ \citenamefont
  {{Demler}}(2015)}]{Grusdt15_2}%
  \BibitemOpen
  \bibfield  {author} {\bibinfo {author} {\bibfnamefont {F.}~\bibnamefont
  {{Grusdt}}}\ and\ \bibinfo {author} {\bibfnamefont {E.}~\bibnamefont
  {{Demler}}},\ }\href@noop {} {\bibfield  {journal} {\bibinfo  {journal}
  {arXiv e-prints}\ ,\ \bibinfo {eid} {arXiv:1510.04934}} (\bibinfo {year}
  {2015})},\ \Eprint {http://arxiv.org/abs/1510.04934} {arXiv:1510.04934
  [cond-mat.quant-gas]} \BibitemShut {NoStop}%
\bibitem [{\citenamefont {Kain}\ and\ \citenamefont {Ling}(2014)}]{Kain14}%
  \BibitemOpen
  \bibfield  {author} {\bibinfo {author} {\bibfnamefont {B.}~\bibnamefont
  {Kain}}\ and\ \bibinfo {author} {\bibfnamefont {H.~Y.}\ \bibnamefont
  {Ling}},\ }\href {\doibase 10.1103/PhysRevA.89.023612} {\bibfield  {journal}
  {\bibinfo  {journal} {Phys. Rev. A}\ }\textbf {\bibinfo {volume} {89}},\
  \bibinfo {pages} {023612} (\bibinfo {year} {2014})}\BibitemShut {NoStop}%
\bibitem [{\citenamefont {Vlietinck}\ \emph {et~al.}(2015)\citenamefont
  {Vlietinck}, \citenamefont {Casteels}, \citenamefont {Van~Houcke},
  \citenamefont {Tempere}, \citenamefont {Ryckebusch},\ and\ \citenamefont
  {Devreese}}]{Vlietinck15}%
  \BibitemOpen
  \bibfield  {author} {\bibinfo {author} {\bibfnamefont {J.}~\bibnamefont
  {Vlietinck}}, \bibinfo {author} {\bibfnamefont {W.}~\bibnamefont {Casteels}},
  \bibinfo {author} {\bibfnamefont {K.}~\bibnamefont {Van~Houcke}}, \bibinfo
  {author} {\bibfnamefont {J.}~\bibnamefont {Tempere}}, \bibinfo {author}
  {\bibfnamefont {J.}~\bibnamefont {Ryckebusch}}, \ and\ \bibinfo {author}
  {\bibfnamefont {J.~T.}\ \bibnamefont {Devreese}},\ }\href@noop {} {\bibfield
  {journal} {\bibinfo  {journal} {New J. Phys.}\ }\textbf {\bibinfo {volume}
  {17}},\ \bibinfo {pages} {033023} (\bibinfo {year} {2015})}\BibitemShut
  {NoStop}%
\bibitem [{\citenamefont {Levinsen}\ \emph {et~al.}(2017)\citenamefont
  {Levinsen}, \citenamefont {Parish}, \citenamefont {Christensen},
  \citenamefont {Arlt},\ and\ \citenamefont {Bruun}}]{Levinsen17}%
  \BibitemOpen
  \bibfield  {author} {\bibinfo {author} {\bibfnamefont {J.}~\bibnamefont
  {Levinsen}}, \bibinfo {author} {\bibfnamefont {M.~M.}\ \bibnamefont
  {Parish}}, \bibinfo {author} {\bibfnamefont {R.~S.}\ \bibnamefont
  {Christensen}}, \bibinfo {author} {\bibfnamefont {J.~J.}\ \bibnamefont
  {Arlt}}, \ and\ \bibinfo {author} {\bibfnamefont {G.~M.}\ \bibnamefont
  {Bruun}},\ }\href {\doibase 10.1103/PhysRevA.96.063622} {\bibfield  {journal}
  {\bibinfo  {journal} {Phys. Rev. A}\ }\textbf {\bibinfo {volume} {96}},\
  \bibinfo {pages} {063622} (\bibinfo {year} {2017})}\BibitemShut {NoStop}%
\bibitem [{\citenamefont {{Lampo}}\ \emph {et~al.}(2018)\citenamefont
  {{Lampo}}, \citenamefont {{Charalambous}}, \citenamefont {{{\'A}ngel
  Garc{\'{\i}}a-March}},\ and\ \citenamefont {{Lewenstein}}}]{Lampo2018}%
  \BibitemOpen
  \bibfield  {author} {\bibinfo {author} {\bibfnamefont {A.}~\bibnamefont
  {{Lampo}}}, \bibinfo {author} {\bibfnamefont {C.}~\bibnamefont
  {{Charalambous}}}, \bibinfo {author} {\bibfnamefont {M.}~\bibnamefont
  {{{\'A}ngel Garc{\'{\i}}a-March}}}, \ and\ \bibinfo {author} {\bibfnamefont
  {M.}~\bibnamefont {{Lewenstein}}},\ }\href@noop {} {\bibfield  {journal}
  {\bibinfo  {journal} {ArXiv:1803.08946}\ } (\bibinfo {year}
  {2018})}\BibitemShut {NoStop}%
\bibitem [{\citenamefont {Nielsen}\ \emph {et~al.}(2019)\citenamefont
  {Nielsen}, \citenamefont {Ardila}, \citenamefont {Bruun},\ and\ \citenamefont
  {Pohl}}]{Nielsen18}%
  \BibitemOpen
  \bibfield  {author} {\bibinfo {author} {\bibfnamefont {K.~K.}\ \bibnamefont
  {Nielsen}}, \bibinfo {author} {\bibfnamefont {L.~A.~P.}\ \bibnamefont
  {Ardila}}, \bibinfo {author} {\bibfnamefont {G.~M.}\ \bibnamefont {Bruun}}, \
  and\ \bibinfo {author} {\bibfnamefont {T.}~\bibnamefont {Pohl}},\ }\href
  {http://iopscience.iop.org/10.1088/1367-2630/ab0a81} {\bibfield  {journal}
  {\bibinfo  {journal} {New Journal of Physics}\ } (\bibinfo {year}
  {2019})}\BibitemShut {NoStop}%
\bibitem [{\citenamefont {Ardila}\ and\ \citenamefont
  {Pohl}(2018)}]{DipolarPolaron18}%
  \BibitemOpen
  \bibfield  {author} {\bibinfo {author} {\bibfnamefont {L.~A.~P.}\
  \bibnamefont {Ardila}}\ and\ \bibinfo {author} {\bibfnamefont
  {T.}~\bibnamefont {Pohl}},\ }\href {\doibase 10.1088/1361-6455/aaf35e}
  {\bibfield  {journal} {\bibinfo  {journal} {Journal of Physics B: Atomic,
  Molecular and Optical Physics}\ }\textbf {\bibinfo {volume} {52}},\ \bibinfo
  {pages} {015004} (\bibinfo {year} {2018})}\BibitemShut {NoStop}%
\bibitem [{\citenamefont {Cucchietti}\ and\ \citenamefont
  {Timmermans}(2006)}]{Cucchietti06}%
  \BibitemOpen
  \bibfield  {author} {\bibinfo {author} {\bibfnamefont {F.~M.}\ \bibnamefont
  {Cucchietti}}\ and\ \bibinfo {author} {\bibfnamefont {E.}~\bibnamefont
  {Timmermans}},\ }\href {\doibase 10.1103/PhysRevLett.96.210401} {\bibfield
  {journal} {\bibinfo  {journal} {Phys. Rev. Lett.}\ }\textbf {\bibinfo
  {volume} {96}},\ \bibinfo {pages} {210401} (\bibinfo {year}
  {2006})}\BibitemShut {NoStop}%
\bibitem [{\citenamefont {Kalas}\ and\ \citenamefont {Blume}(2006)}]{Kalas06}%
  \BibitemOpen
  \bibfield  {author} {\bibinfo {author} {\bibfnamefont {R.~M.}\ \bibnamefont
  {Kalas}}\ and\ \bibinfo {author} {\bibfnamefont {D.}~\bibnamefont {Blume}},\
  }\href {\doibase 10.1103/PhysRevA.73.043608} {\bibfield  {journal} {\bibinfo
  {journal} {Phys. Rev. A}\ }\textbf {\bibinfo {volume} {73}},\ \bibinfo
  {pages} {043608} (\bibinfo {year} {2006})}\BibitemShut {NoStop}%
\bibitem [{\citenamefont {Bruderer}\ \emph {et~al.}(2008)\citenamefont
  {Bruderer}, \citenamefont {Bao},\ and\ \citenamefont {Jaksch}}]{Bruderer08}%
  \BibitemOpen
  \bibfield  {author} {\bibinfo {author} {\bibfnamefont {M.}~\bibnamefont
  {Bruderer}}, \bibinfo {author} {\bibfnamefont {W.}~\bibnamefont {Bao}}, \
  and\ \bibinfo {author} {\bibfnamefont {D.}~\bibnamefont {Jaksch}},\
  }\href@noop {} {\bibfield  {journal} {\bibinfo  {journal} {Eur. Phys. Lett.}\
  }\textbf {\bibinfo {volume} {82}},\ \bibinfo {pages} {30004} (\bibinfo {year}
  {2008})}\BibitemShut {NoStop}%
\bibitem [{\citenamefont {Santamore}\ and\ \citenamefont
  {Timmermans}(2011)}]{Santamore11}%
  \BibitemOpen
  \bibfield  {author} {\bibinfo {author} {\bibfnamefont {D.}~\bibnamefont
  {Santamore}}\ and\ \bibinfo {author} {\bibfnamefont {E.}~\bibnamefont
  {Timmermans}},\ }\href@noop {} {\bibfield  {journal} {\bibinfo  {journal}
  {New J. Phys.}\ }\textbf {\bibinfo {volume} {13}},\ \bibinfo {pages} {103029}
  (\bibinfo {year} {2011})}\BibitemShut {NoStop}%
\bibitem [{\citenamefont {Blinova}\ \emph {et~al.}(2013)\citenamefont
  {Blinova}, \citenamefont {Boshier},\ and\ \citenamefont
  {Timmermans}}]{Blinova13}%
  \BibitemOpen
  \bibfield  {author} {\bibinfo {author} {\bibfnamefont {A.~A.}\ \bibnamefont
  {Blinova}}, \bibinfo {author} {\bibfnamefont {M.~G.}\ \bibnamefont
  {Boshier}}, \ and\ \bibinfo {author} {\bibfnamefont {E.}~\bibnamefont
  {Timmermans}},\ }\href {\doibase 10.1103/PhysRevA.88.053610} {\bibfield
  {journal} {\bibinfo  {journal} {Phys. Rev. A}\ }\textbf {\bibinfo {volume}
  {88}},\ \bibinfo {pages} {053610} (\bibinfo {year} {2013})}\BibitemShut
  {NoStop}%
\bibitem [{\citenamefont {Rath}\ and\ \citenamefont {Schmidt}(2013)}]{Rath13}%
  \BibitemOpen
  \bibfield  {author} {\bibinfo {author} {\bibfnamefont {S.~P.}\ \bibnamefont
  {Rath}}\ and\ \bibinfo {author} {\bibfnamefont {R.}~\bibnamefont {Schmidt}},\
  }\href {\doibase 10.1103/PhysRevA.88.053632} {\bibfield  {journal} {\bibinfo
  {journal} {Phys. Rev. A}\ }\textbf {\bibinfo {volume} {88}},\ \bibinfo
  {pages} {053632} (\bibinfo {year} {2013})}\BibitemShut {NoStop}%
\bibitem [{\citenamefont {Li}\ and\ \citenamefont {Das~Sarma}(2014)}]{Li14}%
  \BibitemOpen
  \bibfield  {author} {\bibinfo {author} {\bibfnamefont {W.}~\bibnamefont
  {Li}}\ and\ \bibinfo {author} {\bibfnamefont {S.}~\bibnamefont {Das~Sarma}},\
  }\href {\doibase 10.1103/PhysRevA.90.013618} {\bibfield  {journal} {\bibinfo
  {journal} {Phys. Rev. A}\ }\textbf {\bibinfo {volume} {90}},\ \bibinfo
  {pages} {013618} (\bibinfo {year} {2014})}\BibitemShut {NoStop}%
\bibitem [{\citenamefont {Christensen}\ \emph {et~al.}(2015)\citenamefont
  {Christensen}, \citenamefont {Levinsen},\ and\ \citenamefont
  {Bruun}}]{Christensen15}%
  \BibitemOpen
  \bibfield  {author} {\bibinfo {author} {\bibfnamefont {R.~S.}\ \bibnamefont
  {Christensen}}, \bibinfo {author} {\bibfnamefont {J.}~\bibnamefont
  {Levinsen}}, \ and\ \bibinfo {author} {\bibfnamefont {G.~M.}\ \bibnamefont
  {Bruun}},\ }\href {\doibase 10.1103/PhysRevLett.115.160401} {\bibfield
  {journal} {\bibinfo  {journal} {Phys. Rev. Lett.}\ }\textbf {\bibinfo
  {volume} {115}},\ \bibinfo {pages} {160401} (\bibinfo {year}
  {2015})}\BibitemShut {NoStop}%
\bibitem [{\citenamefont {Ardila}\ and\ \citenamefont
  {Giorgini}(2015)}]{Ardila15}%
  \BibitemOpen
  \bibfield  {author} {\bibinfo {author} {\bibfnamefont {L.~A.~P.}\
  \bibnamefont {Ardila}}\ and\ \bibinfo {author} {\bibfnamefont
  {S.}~\bibnamefont {Giorgini}},\ }\href {\doibase 10.1103/PhysRevA.92.033612}
  {\bibfield  {journal} {\bibinfo  {journal} {Phys. Rev. A}\ }\textbf {\bibinfo
  {volume} {92}},\ \bibinfo {pages} {033612} (\bibinfo {year}
  {2015})}\BibitemShut {NoStop}%
\bibitem [{\citenamefont {Shchadilova}\ \emph {et~al.}(2016)\citenamefont
  {Shchadilova}, \citenamefont {Schmidt}, \citenamefont {Grusdt},\ and\
  \citenamefont {Demler}}]{Shchadilova16}%
  \BibitemOpen
  \bibfield  {author} {\bibinfo {author} {\bibfnamefont {Y.~E.}\ \bibnamefont
  {Shchadilova}}, \bibinfo {author} {\bibfnamefont {R.}~\bibnamefont
  {Schmidt}}, \bibinfo {author} {\bibfnamefont {F.}~\bibnamefont {Grusdt}}, \
  and\ \bibinfo {author} {\bibfnamefont {E.}~\bibnamefont {Demler}},\ }\href
  {\doibase 10.1103/PhysRevLett.117.113002} {\bibfield  {journal} {\bibinfo
  {journal} {Phys. Rev. Lett.}\ }\textbf {\bibinfo {volume} {117}},\ \bibinfo
  {pages} {113002} (\bibinfo {year} {2016})}\BibitemShut {NoStop}%
\bibitem [{\citenamefont {Grusdt}\ \emph {et~al.}(2017)\citenamefont {Grusdt},
  \citenamefont {Astrakharchik},\ and\ \citenamefont {Demler}}]{Grusdt2017}%
  \BibitemOpen
  \bibfield  {author} {\bibinfo {author} {\bibfnamefont {F.}~\bibnamefont
  {Grusdt}}, \bibinfo {author} {\bibfnamefont {G.~E.}\ \bibnamefont
  {Astrakharchik}}, \ and\ \bibinfo {author} {\bibfnamefont {E.}~\bibnamefont
  {Demler}},\ }\href {\doibase 10.1088/1367-2630/aa8a2e} {\bibfield  {journal}
  {\bibinfo  {journal} {New Journal of Physics}\ }\textbf {\bibinfo {volume}
  {19}},\ \bibinfo {pages} {103035} (\bibinfo {year} {2017})}\BibitemShut
  {NoStop}%
\bibitem [{\citenamefont {Kain}\ and\ \citenamefont {Ling}(2018)}]{Kain18}%
  \BibitemOpen
  \bibfield  {author} {\bibinfo {author} {\bibfnamefont {B.}~\bibnamefont
  {Kain}}\ and\ \bibinfo {author} {\bibfnamefont {H.~Y.}\ \bibnamefont
  {Ling}},\ }\href {\doibase 10.1103/PhysRevA.98.033610} {\bibfield  {journal}
  {\bibinfo  {journal} {Phys. Rev. A}\ }\textbf {\bibinfo {volume} {98}},\
  \bibinfo {pages} {033610} (\bibinfo {year} {2018})}\BibitemShut {NoStop}%
\bibitem [{\citenamefont {Boudjem\^aa}(2014)}]{Abdelaali14}%
  \BibitemOpen
  \bibfield  {author} {\bibinfo {author} {\bibfnamefont {A.}~\bibnamefont
  {Boudjem\^aa}},\ }\href {\doibase 10.1103/PhysRevA.90.013628} {\bibfield
  {journal} {\bibinfo  {journal} {Phys. Rev. A}\ }\textbf {\bibinfo {volume}
  {90}},\ \bibinfo {pages} {013628} (\bibinfo {year} {2014})}\BibitemShut
  {NoStop}%
\bibitem [{\citenamefont {Boudjem{\^{a}}a}(2014)}]{Abdelaali14-2}%
  \BibitemOpen
  \bibfield  {author} {\bibinfo {author} {\bibfnamefont {A.}~\bibnamefont
  {Boudjem{\^{a}}a}},\ }\href {\doibase 10.1088/1751-8113/48/4/045002}
  {\bibfield  {journal} {\bibinfo  {journal} {Journal of Physics A:
  Mathematical and Theoretical}\ }\textbf {\bibinfo {volume} {48}},\ \bibinfo
  {pages} {045002} (\bibinfo {year} {2014})}\BibitemShut {NoStop}%
\bibitem [{\citenamefont {{Mistakidis}}\ \emph {et~al.}(2018)\citenamefont
  {{Mistakidis}}, \citenamefont {{Volosniev}}, \citenamefont {{Zinner}},\ and\
  \citenamefont {{Schmelcher}}}]{Mistakidis18}%
  \BibitemOpen
  \bibfield  {author} {\bibinfo {author} {\bibfnamefont {S.~I.}\ \bibnamefont
  {{Mistakidis}}}, \bibinfo {author} {\bibfnamefont {A.~G.}\ \bibnamefont
  {{Volosniev}}}, \bibinfo {author} {\bibfnamefont {N.~T.}\ \bibnamefont
  {{Zinner}}}, \ and\ \bibinfo {author} {\bibfnamefont {P.}~\bibnamefont
  {{Schmelcher}}},\ }\href@noop {} {\bibfield  {journal} {\bibinfo  {journal}
  {arXiv e-prints}\ ,\ \bibinfo {eid} {arXiv:1809.01889}} (\bibinfo {year}
  {2018})},\ \Eprint {http://arxiv.org/abs/1809.01889} {arXiv:1809.01889
  [cond-mat.quant-gas]} \BibitemShut {NoStop}%
\bibitem [{\citenamefont {Mistakidis}\ \emph {et~al.}(2019)\citenamefont
  {Mistakidis}, \citenamefont {Katsimiga}, \citenamefont {Koutentakis},
  \citenamefont {Busch},\ and\ \citenamefont {Schmelcher}}]{Mistakidis19}%
  \BibitemOpen
  \bibfield  {author} {\bibinfo {author} {\bibfnamefont {S.~I.}\ \bibnamefont
  {Mistakidis}}, \bibinfo {author} {\bibfnamefont {G.~C.}\ \bibnamefont
  {Katsimiga}}, \bibinfo {author} {\bibfnamefont {G.~M.}\ \bibnamefont
  {Koutentakis}}, \bibinfo {author} {\bibfnamefont {T.}~\bibnamefont {Busch}},
  \ and\ \bibinfo {author} {\bibfnamefont {P.}~\bibnamefont {Schmelcher}},\
  }\href {\doibase 10.1103/PhysRevLett.122.183001} {\bibfield  {journal}
  {\bibinfo  {journal} {Phys. Rev. Lett.}\ }\textbf {\bibinfo {volume} {122}},\
  \bibinfo {pages} {183001} (\bibinfo {year} {2019})}\BibitemShut {NoStop}%
\bibitem [{\citenamefont {Drescher}\ \emph {et~al.}(2019)\citenamefont
  {Drescher}, \citenamefont {Salmhofer},\ and\ \citenamefont
  {Enss}}]{Drescher18}%
  \BibitemOpen
  \bibfield  {author} {\bibinfo {author} {\bibfnamefont {M.}~\bibnamefont
  {Drescher}}, \bibinfo {author} {\bibfnamefont {M.}~\bibnamefont {Salmhofer}},
  \ and\ \bibinfo {author} {\bibfnamefont {T.}~\bibnamefont {Enss}},\ }\href
  {\doibase 10.1103/PhysRevA.99.023601} {\bibfield  {journal} {\bibinfo
  {journal} {Phys. Rev. A}\ }\textbf {\bibinfo {volume} {99}},\ \bibinfo
  {pages} {023601} (\bibinfo {year} {2019})}\BibitemShut {NoStop}%
\bibitem [{\citenamefont {Guenther}\ \emph {et~al.}(2018)\citenamefont
  {Guenther}, \citenamefont {Massignan}, \citenamefont {Lewenstein},\ and\
  \citenamefont {Bruun}}]{Guenther18}%
  \BibitemOpen
  \bibfield  {author} {\bibinfo {author} {\bibfnamefont {N.-E.}\ \bibnamefont
  {Guenther}}, \bibinfo {author} {\bibfnamefont {P.}~\bibnamefont {Massignan}},
  \bibinfo {author} {\bibfnamefont {M.}~\bibnamefont {Lewenstein}}, \ and\
  \bibinfo {author} {\bibfnamefont {G.~M.}\ \bibnamefont {Bruun}},\ }\href
  {\doibase 10.1103/PhysRevLett.120.050405} {\bibfield  {journal} {\bibinfo
  {journal} {Phys. Rev. Lett.}\ }\textbf {\bibinfo {volume} {120}},\ \bibinfo
  {pages} {050405} (\bibinfo {year} {2018})}\BibitemShut {NoStop}%
\bibitem [{\citenamefont {Levinsen}\ \emph {et~al.}(2015)\citenamefont
  {Levinsen}, \citenamefont {Parish},\ and\ \citenamefont
  {Bruun}}]{Levinsen15}%
  \BibitemOpen
  \bibfield  {author} {\bibinfo {author} {\bibfnamefont {J.}~\bibnamefont
  {Levinsen}}, \bibinfo {author} {\bibfnamefont {M.~M.}\ \bibnamefont
  {Parish}}, \ and\ \bibinfo {author} {\bibfnamefont {G.~M.}\ \bibnamefont
  {Bruun}},\ }\href {\doibase 10.1103/PhysRevLett.115.125302} {\bibfield
  {journal} {\bibinfo  {journal} {Phys. Rev. Lett.}\ }\textbf {\bibinfo
  {volume} {115}},\ \bibinfo {pages} {125302} (\bibinfo {year}
  {2015})}\BibitemShut {NoStop}%
\bibitem [{\citenamefont {Sun}\ \emph {et~al.}(2017)\citenamefont {Sun},
  \citenamefont {Zhai},\ and\ \citenamefont {Cui}}]{Sun17}%
  \BibitemOpen
  \bibfield  {author} {\bibinfo {author} {\bibfnamefont {M.}~\bibnamefont
  {Sun}}, \bibinfo {author} {\bibfnamefont {H.}~\bibnamefont {Zhai}}, \ and\
  \bibinfo {author} {\bibfnamefont {X.}~\bibnamefont {Cui}},\ }\href {\doibase
  10.1103/PhysRevLett.119.013401} {\bibfield  {journal} {\bibinfo  {journal}
  {Phys. Rev. Lett.}\ }\textbf {\bibinfo {volume} {119}},\ \bibinfo {pages}
  {013401} (\bibinfo {year} {2017})}\BibitemShut {NoStop}%
\bibitem [{\citenamefont {Dehkharghani}\ \emph {et~al.}(2018)\citenamefont
  {Dehkharghani}, \citenamefont {Volosniev},\ and\ \citenamefont
  {Zinner}}]{Dehkharghani18}%
  \BibitemOpen
  \bibfield  {author} {\bibinfo {author} {\bibfnamefont {A.~S.}\ \bibnamefont
  {Dehkharghani}}, \bibinfo {author} {\bibfnamefont {A.~G.}\ \bibnamefont
  {Volosniev}}, \ and\ \bibinfo {author} {\bibfnamefont {N.~T.}\ \bibnamefont
  {Zinner}},\ }\href {\doibase 10.1103/PhysRevLett.121.080405} {\bibfield
  {journal} {\bibinfo  {journal} {Phys. Rev. Lett.}\ }\textbf {\bibinfo
  {volume} {121}},\ \bibinfo {pages} {080405} (\bibinfo {year}
  {2018})}\BibitemShut {NoStop}%
\bibitem [{\citenamefont {Camacho-Guardian}\ \emph {et~al.}(2018)\citenamefont
  {Camacho-Guardian}, \citenamefont {Pe\~na Ardila}, \citenamefont {Pohl},\
  and\ \citenamefont {Bruun}}]{Camacho2018b}%
  \BibitemOpen
  \bibfield  {author} {\bibinfo {author} {\bibfnamefont {A.}~\bibnamefont
  {Camacho-Guardian}}, \bibinfo {author} {\bibfnamefont {L.~A.}\ \bibnamefont
  {Pe\~na Ardila}}, \bibinfo {author} {\bibfnamefont {T.}~\bibnamefont {Pohl}},
  \ and\ \bibinfo {author} {\bibfnamefont {G.~M.}\ \bibnamefont {Bruun}},\
  }\href {\doibase 10.1103/PhysRevLett.121.013401} {\bibfield  {journal}
  {\bibinfo  {journal} {Phys. Rev. Lett.}\ }\textbf {\bibinfo {volume} {121}},\
  \bibinfo {pages} {013401} (\bibinfo {year} {2018})}\BibitemShut {NoStop}%
\bibitem [{\citenamefont {Ardila}\ and\ \citenamefont
  {Giorgini}(2016)}]{Ardila16}%
  \BibitemOpen
  \bibfield  {author} {\bibinfo {author} {\bibfnamefont {L.~A.~P.}\
  \bibnamefont {Ardila}}\ and\ \bibinfo {author} {\bibfnamefont
  {S.}~\bibnamefont {Giorgini}},\ }\href {\doibase 10.1103/PhysRevA.94.063640}
  {\bibfield  {journal} {\bibinfo  {journal} {Phys. Rev. A}\ }\textbf {\bibinfo
  {volume} {94}},\ \bibinfo {pages} {063640} (\bibinfo {year}
  {2016})}\BibitemShut {NoStop}%
\bibitem [{\citenamefont {Parisi}\ and\ \citenamefont
  {Giorgini}(2017)}]{Parisi16}%
  \BibitemOpen
  \bibfield  {author} {\bibinfo {author} {\bibfnamefont {L.}~\bibnamefont
  {Parisi}}\ and\ \bibinfo {author} {\bibfnamefont {S.}~\bibnamefont
  {Giorgini}},\ }\href {\doibase 10.1103/PhysRevA.95.023619} {\bibfield
  {journal} {\bibinfo  {journal} {Phys. Rev. A}\ }\textbf {\bibinfo {volume}
  {95}},\ \bibinfo {pages} {023619} (\bibinfo {year} {2017})}\BibitemShut
  {NoStop}%
\bibitem [{\citenamefont {Massignan}\ \emph {et~al.}(2014)\citenamefont
  {Massignan}, \citenamefont {Zaccanti},\ and\ \citenamefont
  {Bruun}}]{Massignan14}%
  \BibitemOpen
  \bibfield  {author} {\bibinfo {author} {\bibfnamefont {P.}~\bibnamefont
  {Massignan}}, \bibinfo {author} {\bibfnamefont {M.}~\bibnamefont {Zaccanti}},
  \ and\ \bibinfo {author} {\bibfnamefont {G.~M.}\ \bibnamefont {Bruun}},\
  }\href@noop {} {\bibfield  {journal} {\bibinfo  {journal} {Rep. Prog. Phys.}\
  }\textbf {\bibinfo {volume} {77}},\ \bibinfo {pages} {034401} (\bibinfo
  {year} {2014})}\BibitemShut {NoStop}%
\bibitem [{\citenamefont {Ngampruetikorn}\ \emph {et~al.}(2012)\citenamefont
  {Ngampruetikorn}, \citenamefont {Levinsen},\ and\ \citenamefont
  {Parish}}]{Nga12}%
  \BibitemOpen
  \bibfield  {author} {\bibinfo {author} {\bibfnamefont {V.}~\bibnamefont
  {Ngampruetikorn}}, \bibinfo {author} {\bibfnamefont {J.}~\bibnamefont
  {Levinsen}}, \ and\ \bibinfo {author} {\bibfnamefont {M.~M.}\ \bibnamefont
  {Parish}},\ }\href {\doibase 10.1209/0295-5075/98/30005} {\bibfield
  {journal} {\bibinfo  {journal} {{EPL} (Europhysics Letters)}\ }\textbf
  {\bibinfo {volume} {98}},\ \bibinfo {pages} {30005} (\bibinfo {year}
  {2012})}\BibitemShut {NoStop}%
\bibitem [{\citenamefont {Ngampruetikorn}\ \emph {et~al.}(2013)\citenamefont
  {Ngampruetikorn}, \citenamefont {Parish},\ and\ \citenamefont
  {Levinsen}}]{Nga13}%
  \BibitemOpen
  \bibfield  {author} {\bibinfo {author} {\bibfnamefont {V.}~\bibnamefont
  {Ngampruetikorn}}, \bibinfo {author} {\bibfnamefont {M.~M.}\ \bibnamefont
  {Parish}}, \ and\ \bibinfo {author} {\bibfnamefont {J.}~\bibnamefont
  {Levinsen}},\ }\href {\doibase 10.1209/0295-5075/102/13001} {\bibfield
  {journal} {\bibinfo  {journal} {{EPL} (Europhysics Letters)}\ }\textbf
  {\bibinfo {volume} {102}},\ \bibinfo {pages} {13001} (\bibinfo {year}
  {2013})}\BibitemShut {NoStop}%
\bibitem [{\citenamefont {Schmidt}\ \emph {et~al.}(2012)\citenamefont
  {Schmidt}, \citenamefont {Enss}, \citenamefont {Pietil\"a},\ and\
  \citenamefont {Demler}}]{SchmidtFermi}%
  \BibitemOpen
  \bibfield  {author} {\bibinfo {author} {\bibfnamefont {R.}~\bibnamefont
  {Schmidt}}, \bibinfo {author} {\bibfnamefont {T.}~\bibnamefont {Enss}},
  \bibinfo {author} {\bibfnamefont {V.}~\bibnamefont {Pietil\"a}}, \ and\
  \bibinfo {author} {\bibfnamefont {E.}~\bibnamefont {Demler}},\ }\href
  {\doibase 10.1103/PhysRevA.85.021602} {\bibfield  {journal} {\bibinfo
  {journal} {Phys. Rev. A}\ }\textbf {\bibinfo {volume} {85}},\ \bibinfo
  {pages} {021602} (\bibinfo {year} {2012})}\BibitemShut {NoStop}%
\bibitem [{\citenamefont {Sidler}\ \emph {et~al.}(2017)\citenamefont {Sidler},
  \citenamefont {Back}, \citenamefont {Cotlet}, \citenamefont {Srivastava},
  \citenamefont {Fink}, \citenamefont {Kroner}, \citenamefont {Demler},\ and\
  \citenamefont {Imamoglu}}]{Sidler17}%
  \BibitemOpen
  \bibfield  {author} {\bibinfo {author} {\bibfnamefont {M.}~\bibnamefont
  {Sidler}}, \bibinfo {author} {\bibfnamefont {P.}~\bibnamefont {Back}},
  \bibinfo {author} {\bibfnamefont {O.}~\bibnamefont {Cotlet}}, \bibinfo
  {author} {\bibfnamefont {A.}~\bibnamefont {Srivastava}}, \bibinfo {author}
  {\bibfnamefont {T.}~\bibnamefont {Fink}}, \bibinfo {author} {\bibfnamefont
  {M.}~\bibnamefont {Kroner}}, \bibinfo {author} {\bibfnamefont
  {E.}~\bibnamefont {Demler}}, \ and\ \bibinfo {author} {\bibfnamefont
  {A.}~\bibnamefont {Imamoglu}},\ }\href@noop {} {\bibfield  {journal}
  {\bibinfo  {journal} {Nat. Phys.}\ }\textbf {\bibinfo {volume} {13}},\
  \bibinfo {pages} {255} (\bibinfo {year} {2017})}\BibitemShut {NoStop}%
\bibitem [{\citenamefont {Casteels}\ \emph {et~al.}(2012)\citenamefont
  {Casteels}, \citenamefont {Tempere},\ and\ \citenamefont
  {Devreese}}]{Casteels12}%
  \BibitemOpen
  \bibfield  {author} {\bibinfo {author} {\bibfnamefont {W.}~\bibnamefont
  {Casteels}}, \bibinfo {author} {\bibfnamefont {J.}~\bibnamefont {Tempere}}, \
  and\ \bibinfo {author} {\bibfnamefont {J.~T.}\ \bibnamefont {Devreese}},\
  }\href {\doibase 10.1103/PhysRevA.86.043614} {\bibfield  {journal} {\bibinfo
  {journal} {Phys. Rev. A}\ }\textbf {\bibinfo {volume} {86}},\ \bibinfo
  {pages} {043614} (\bibinfo {year} {2012})}\BibitemShut {NoStop}%
\bibitem [{\citenamefont {Grusdt}\ and\ \citenamefont
  {Fleischhauer}(2016)}]{Grusdt16}%
  \BibitemOpen
  \bibfield  {author} {\bibinfo {author} {\bibfnamefont {F.}~\bibnamefont
  {Grusdt}}\ and\ \bibinfo {author} {\bibfnamefont {M.}~\bibnamefont
  {Fleischhauer}},\ }\href@noop {} {\bibfield  {journal} {\bibinfo  {journal}
  {Phys. Rev. Lett.}\ }\textbf {\bibinfo {volume} {116}},\ \bibinfo {pages}
  {053602} (\bibinfo {year} {2016})}\BibitemShut {NoStop}%
\bibitem [{\citenamefont {Pastukhov}(2018)}]{Pastukhov18}%
  \BibitemOpen
  \bibfield  {author} {\bibinfo {author} {\bibfnamefont {V.}~\bibnamefont
  {Pastukhov}},\ }\href@noop {} {\bibfield  {journal} {\bibinfo  {journal} {J.
  Phys. B: At. Mol. Opt. Phys.}\ }\textbf {\bibinfo {volume} {51}},\ \bibinfo
  {pages} {155203} (\bibinfo {year} {2018})}\BibitemShut {NoStop}%
\bibitem [{\citenamefont {Ardila}(2015)}]{ThesisPOL}%
  \BibitemOpen
  \bibfield  {author} {\bibinfo {author} {\bibfnamefont {L.~A.~P.}\
  \bibnamefont {Ardila}},\ }\emph {\bibinfo {title} {Impurities in a
  Bose-Einstein condensate using quantum Monte-Carlo methods: ground-state
  properties.}},\ \href@noop {} {Ph.D. thesis},\ \bibinfo  {school} {University
  of Trento} (\bibinfo {year} {2015})\BibitemShut {NoStop}%
\bibitem [{\citenamefont {Ville}\ \emph {et~al.}(2018)\citenamefont {Ville},
  \citenamefont {Saint-Jalm}, \citenamefont {Le~Cerf}, \citenamefont
  {Aidelsburger}, \citenamefont {Nascimb\`ene}, \citenamefont {Dalibard},\ and\
  \citenamefont {Beugnon}}]{Ville18}%
  \BibitemOpen
  \bibfield  {author} {\bibinfo {author} {\bibfnamefont {J.~L.}\ \bibnamefont
  {Ville}}, \bibinfo {author} {\bibfnamefont {R.}~\bibnamefont {Saint-Jalm}},
  \bibinfo {author} {\bibfnamefont {E.}~\bibnamefont {Le~Cerf}}, \bibinfo
  {author} {\bibfnamefont {M.}~\bibnamefont {Aidelsburger}}, \bibinfo {author}
  {\bibfnamefont {S.}~\bibnamefont {Nascimb\`ene}}, \bibinfo {author}
  {\bibfnamefont {J.}~\bibnamefont {Dalibard}}, \ and\ \bibinfo {author}
  {\bibfnamefont {J.}~\bibnamefont {Beugnon}},\ }\href {\doibase
  10.1103/PhysRevLett.121.145301} {\bibfield  {journal} {\bibinfo  {journal}
  {Phys. Rev. Lett.}\ }\textbf {\bibinfo {volume} {121}},\ \bibinfo {pages}
  {145301} (\bibinfo {year} {2018})}\BibitemShut {NoStop}%
\bibitem [{not({\natexlab{a}})}]{note1}%
  \BibitemOpen
  \href@noop {} {}\bibinfo {note} {Notice that results with $Z_0\simeq0$ are
  obtained from variational calculations employing the wavefunction which
  interpolates between the attractive and the repulsive branch.}\BibitemShut
  {Stop}%
\bibitem [{\citenamefont {Reatto}\ and\ \citenamefont
  {Chester}(1967)}]{ReattoChester67}%
  \BibitemOpen
  \bibfield  {author} {\bibinfo {author} {\bibfnamefont {L.}~\bibnamefont
  {Reatto}}\ and\ \bibinfo {author} {\bibfnamefont {G.~V.}\ \bibnamefont
  {Chester}},\ }\href {\doibase 10.1103/PhysRev.155.88} {\bibfield  {journal}
  {\bibinfo  {journal} {Phys. Rev.}\ }\textbf {\bibinfo {volume} {155}},\
  \bibinfo {pages} {88} (\bibinfo {year} {1967})}\BibitemShut {NoStop}%
\bibitem [{not({\natexlab{b}})}]{note}%
  \BibitemOpen
  \href@noop {} {}\bibinfo {note} {We expect perturbation theory for $m/m^\ast$
  and $Z_0$ to be valid when $\ln(1/na_B^2)\ll\ln(1/na^2)$ which, due to the
  small value of $na_B^2$, requires coupling strengths $|\ln(k_Fa)|$ much
  larger than the ones shown in Figs.~3-4.}\BibitemShut {Stop}%
\bibitem [{\citenamefont {Chien}\ \emph {et~al.}(2015)\citenamefont {Chien},
  \citenamefont {Peotta},\ and\ \citenamefont {Di~Ventra}}]{Chien15}%
  \BibitemOpen
  \bibfield  {author} {\bibinfo {author} {\bibfnamefont {C.}~\bibnamefont
  {Chien}}, \bibinfo {author} {\bibfnamefont {S.}~\bibnamefont {Peotta}}, \
  and\ \bibinfo {author} {\bibfnamefont {M.}~\bibnamefont {Di~Ventra}},\ }\href
  {\doibase https://doi.org/10.1038/nphys3531} {\bibfield  {journal} {\bibinfo
  {journal} {Nature Phys}\ }\textbf {\bibinfo {volume} {11}},\ \bibinfo {pages}
  {998–1004} (\bibinfo {year} {2015})}\BibitemShut {NoStop}%
\bibitem [{\citenamefont {Krinner}\ \emph {et~al.}(2017)\citenamefont
  {Krinner}, \citenamefont {Esslinger},\ and\ \citenamefont
  {Brantut}}]{Krinner17}%
  \BibitemOpen
  \bibfield  {author} {\bibinfo {author} {\bibfnamefont {S.}~\bibnamefont
  {Krinner}}, \bibinfo {author} {\bibfnamefont {T.}~\bibnamefont {Esslinger}},
  \ and\ \bibinfo {author} {\bibfnamefont {J.-P.}\ \bibnamefont {Brantut}},\
  }\href {\doibase 10.1088/1361-648x/aa74a1} {\bibfield  {journal} {\bibinfo
  {journal} {Journal of Physics: Condensed Matter}\ }\textbf {\bibinfo {volume}
  {29}},\ \bibinfo {pages} {343003} (\bibinfo {year} {2017})}\BibitemShut
  {NoStop}%
\bibitem [{\citenamefont {Astrakharchik}\ and\ \citenamefont
  {Pitaevskii}(2004)}]{Astrakharchik04}%
  \BibitemOpen
  \bibfield  {author} {\bibinfo {author} {\bibfnamefont {G.~E.}\ \bibnamefont
  {Astrakharchik}}\ and\ \bibinfo {author} {\bibfnamefont {L.~P.}\ \bibnamefont
  {Pitaevskii}},\ }\href {\doibase 10.1103/PhysRevA.70.013608} {\bibfield
  {journal} {\bibinfo  {journal} {Phys. Rev. A}\ }\textbf {\bibinfo {volume}
  {70}},\ \bibinfo {pages} {013608} (\bibinfo {year} {2004})}\BibitemShut
  {NoStop}%
\bibitem [{\citenamefont {{Astrakharchik}}(2014)}]{AstrakharchikPhD}%
  \BibitemOpen
  \bibfield  {author} {\bibinfo {author} {\bibfnamefont {G.~E.}\ \bibnamefont
  {{Astrakharchik}}},\ }\href@noop {} {\bibfield  {journal} {\bibinfo
  {journal} {arXiv e-prints}\ ,\ \bibinfo {eid} {arXiv:1412.4529}} (\bibinfo
  {year} {2014})},\ \Eprint {http://arxiv.org/abs/1412.4529} {arXiv:1412.4529
  [cond-mat.quant-gas]} \BibitemShut {NoStop}%
\end{thebibliography}%

\end{document}